\documentclass[aps,prd,superscriptaddress,showpacs,showkeys]{revtex4-1}  
\usepackage{graphicx,color} 
\usepackage{xcolor}
\usepackage{rotating}
\usepackage{amssymb}
\usepackage{amsfonts}
\usepackage{amsmath}
\usepackage{xspace}
\usepackage{mathtools} 
\usepackage{verbatim}
\DeclarePairedDelimiter\bra{\langle}{\rvert}
\DeclarePairedDelimiter\ket{\lvert}{\rangle}
\DeclarePairedDelimiter\braket{\langle}{\rangle}
\DeclarePairedDelimiterX\Braket[2]{\langle}{\rangle}{#1 \delimsize\vert #2}
\usepackage{dsfont}
\usepackage{slashed}
\newcommand\hatslashed[1]{{\hat{#1}\mathllap{\slashed{#1}}}}
\usepackage{caption}
\usepackage{ragged2e}
\usepackage[hidelinks]{hyperref}

\usepackage{listings}
\makeindex

\usepackage[thinlines]{easytable}

\usepackage[super]{nth}

\newcommand{\be}{\begin{equation}}
\newcommand{\ee}{\end{equation}}
\newcommand{\bea}{\begin{eqnarray}}
\newcommand{\eea}{\end{eqnarray}}
\newcommand{\reals}{{\rm I\!R}}

\newcommand\ie{\mbox{\textit{i.\,e.}}\xspace}
\newcommand\cf{\mbox{c.\,f.}\xspace}
\newcommand\eg{\mbox{e.\,g.}\xspace}

\newcommand\D{\mathrm{d}}

\newcommand\Ord{\mathcal{O}}

\newcommand\hil{\mathcal{H}}
\newcommand{\var}[1]{\left(\sigma^2_{#1}\right)}

\begin{document}

\title{Relativistic Extended Uncertainty Principle from Spacetime Curvature}




\author{Fabian Wagner}
\email[]{fabian.wagner@usz.edu.pl}
\affiliation{Institute of Physics, University of Szczecin, Wielkopolska 15, 70-451 Szczecin, Poland}
\date{\today}
\begin{abstract}
The investigations presented in this study are directed at relativistic modifications of the uncertainty relation derived from the curvature of the background spacetime. These findings generalize previous work which is recovered in the nonrelativistic limit. Applying the 3+1-splitting in accordance with the ADM-formalism, we find the relativistic physical momentum operator and compute its standard deviation for wave functions confined to a geodesic ball on a spacelike hypersurface. Its radius can then be understood as a measure of position uncertainty. Under the assumtion of small position uncertainties in comparison to background curvature length scales, we obtain the corresponding corrections to the uncertainty relation in flat space. Those depend on the Ricci scalar of the effective spatial metric, the particle is moving on, and, if there are nonvanishing time-space components of the spacetime metric, gradients of the shift vector and the lapse function. Interestingly, this result is applicable not only to massive but also to massless particles. Over all, this is not a covariant, yet a consistently general relativistic approach. We further speculate on a possible covariant extension.
\end{abstract}

\pacs{}
\keywords{}

\maketitle

\section{Introduction}
\label{intro}

Deformations of the Heisenberg algebra reflecting the influence of classical and quantum gravity on nonrelativistic quantum mechanics \cite{Mead64,Mead66,Padmanabhan87,Ng93,Maggiore93a,Amelino-Camelia94,Garay94,Adler99a,Scardigli99,Capozziello99,Camacho02,Calmet04,Ghosh10,Casadio13,Casadio15}, as popularized by findings in string theory \cite{Amati87,Gross87a,Gross87b,Amati88,Konishi89}, are consistently gaining in importance in the community of quantum gravity phenomenology. They imply modifications of the uncertainty relation commonly known as generalized \cite{Maggiore93c,Kempf94,Kempf96a,Benczik02,Das12,Petruzziello20,Buoninfante20} and extended uncertainty principles (EUPs) \cite{Bambi08a,Mignemi09,Ghosh09,CostaFilho16} given, that they are momentum or position-dependent, respectively. This peculiar behaviour can be inferred from the Robertson relation \cite{Robertson29,Schroedinger30} linking the standard deviations of the position and momentum operators to their commutator
\begin{equation}
    \sigma_p\sigma_x\geq\frac{\left|\braket*{\hat{x},\hat{p}}\right|}{2}.
\end{equation}
In one-dimensional quantum mechanics, for example, an EUP may be obtained from the algebra of observables
\begin{equation}
    [\hat{x},\hat{p}]=i\hbar\left(1+\frac{p_{\text{min}}^2\hat{x}^2}{4\hbar^2}\right).
\end{equation}
Then, the resulting theory implies a restricted resolution of momentum measurements $\sigma_p>p_{\text{min}}$ or a maximal wave length, akin to the temperature of spacetimes containing cosmological horizons. Similarly, momentum dependent deformations often imply a minimum length. Therefore, EUPs are supposed to be vaguely related to the curvature of the background spacetime \cite{Mignemi09,Ghosh09}, while GUPs may be understood as alternative description of quantum mechanics on curved momentum space \cite{Chang10,Wagner21,Singh21}.

This loose connection was put on firm ground recently \cite{Schuermann18,Dabrowski19,Dabrowski20} as EUPs where derived from the sole assumption that the theory is staged on a curved three-dimensional geometry. This was further generalized to curved spacetime in Ref. \cite{Petruzziello21}. Those studies as well as most of the work on GUPs and EUPs have in common, that they are restricted to the nonrelativistic context (some notable exceptions are provided in Refs. \cite{Todorinov18,Pye19,Gine21,Singh21}). The present paper is intended to relax this assumption by dropping the nonrelativistic limit altogether and thus obtain an uncertainty relation which can be applied to fast as well as massless particles in a general relativistic setting.

In that vein, we allow the four-dimensional background manifold describing position space to be nontrivial. As a result, the relativistic physical three-momentum $p_i$ of a particle moving on this background, which is the observable of interest, does not equal the canonical conjugate $\pi_i$ to the positions $x^i.$ Its explicit form is obtained given the ADM-splitting \cite{Arnowitt60,Arnowitt08}. Turning to the quantum theory, the algebra of canonical observables on spacelike hypersurfaces is assumed to be unaltered
\begin{equation}
    [\hat{x}^i,\hat{x}^j]=0\hspace{1cm}[\hat{\pi}_i,\hat{\pi}_j]=0\hspace{1cm}[\hat{x}^i,\hat{\pi}_j]=i\hbar\delta^i_j\label{Heisalg}
\end{equation}
where Latin indices describe spacelike coordinates while Greek ones indicate a description of spacetime. By analogy with the classical case, the physical relativistic momentum operator transcends the canonical one. In particular, it is plagued by ordering ambiguities. However, we show, that the central finding of this paper is independent of ordering. 

The algebra \eqref{Heisalg} implies, that there are no modifications to Heisenberg's uncertainty principle through the Robertson relation. However, this inequality is not the only way an uncertainty relation in quantum mechanics may be formulated \cite{Maassen88,Maccone14,Coles17}. In principle, the rather vague motivations behind the EUPs cannot be deployed as means of distinguishing between those different approaches. To the contrary, on the GUP-side there are a number of alternative approaches towards a minimum length, for example, by superposition of geometries \cite{Lake18,Lake20} or direct inclusion into differential geometry \cite{Padmanabhan96,Kothawala13,Padmanabhan15}. For the following considerations we rely on the recently found alternative mentioned above, which has a rather operational interpretation and is straight-forwardly generalizable to curved manifolds \cite{Schuermann09,Schuermann18,Dabrowski19,Dabrowski20,Petruzziello21}. The main idea behind this relation consists in confining the theory to a compact domain, in this case a geodesic ball. Then, it is possible to interpret a diffeomorphism invariant measure of its size, here the radius of the ball, as the position uncertainty. In this setting, the standard deviation of the momentum operator develops a global minimum which is dependent on that very measure of uncertainty thus yielding the desired relation.

Assuming the position uncertainty to be small in comparison to background curvature length scales, the effective spatial metric, the particle is subjected to, can be approximated in terms of Riemann normal coordinates. This allows for the perturbative derivation of the uncertainty relation which is obtained to quadratic order in the radius of the ball. 

The paper is organized as follows. First, we introduce the general idea behind this type of uncertainty relation in section \ref{sec:uncrel}. Section \ref{sec:momop} is aimed at deriving the relativistic physical momentum operator. The corrections to the uncertainty relation in flat spacetime are computed in section \ref{sec:explicitsol}, while section \ref{sec:disc} summarizes the conclusions drawn from this result.

\section{Uncertainty relation\label{sec:uncrel}}

This section may be understood as an introduction to the alternative to the Robertson relation we alluded to in the preceding section. The formalism, which is elaborated upon in the present section, was introduced in Refs. \cite{Schuermann09,Schuermann18} and further expanded upon by the author and collaborators in Refs. \cite{Dabrowski19,Dabrowski20,Petruzziello21}. It is instructive to consider it first in the flat case from which the generalization to a curved background is straight-forward.

\subsection{Standard deviation of the momentum operator}\label{subsec:momunc}

Assume that the effective line element on hypersurfaces of constant time reads
\begin{equation}
    \D s^2=\delta_{ij}\D x^i\D x^j
\end{equation}
featuring the Kronecker delta $\delta_{ij}.$ Further asserting the canonical commutation relations \eqref{Heisalg} to be satisfied, the position space representation of the momentum operator may be given in terms of partial derivatives
\begin{equation}
    \hat{\pi}_a\psi=-i\hbar\partial_a\psi
\end{equation}
with a general position space-wave function $\psi.$ Throughout this paper, we will time and again come back to calculating the standard deviation of the momentum operator $\hat{p}$ which, assuming $p_a=\pi_a$ as usual for a nonrelativistic particle in a flat background (more on this below), reads
\begin{equation}
    \sigma_p=\sqrt{\braket*{\hat{\pi}^2}-\delta^{ab}\braket*{\hat{\pi}_a}\braket*{\hat{\pi}_b}}.\label{flatstanddev}
\end{equation}
This quantity will be the measure of momentum uncertainty used in this work. The position uncertainty, however, will be represented differently.

\subsection{Position uncertainty as size of a compact domain}\label{subsec:posunc}

\begin{figure}
\begin{minipage}{.48\textwidth}
    \centering
    \includegraphics[width=\linewidth]{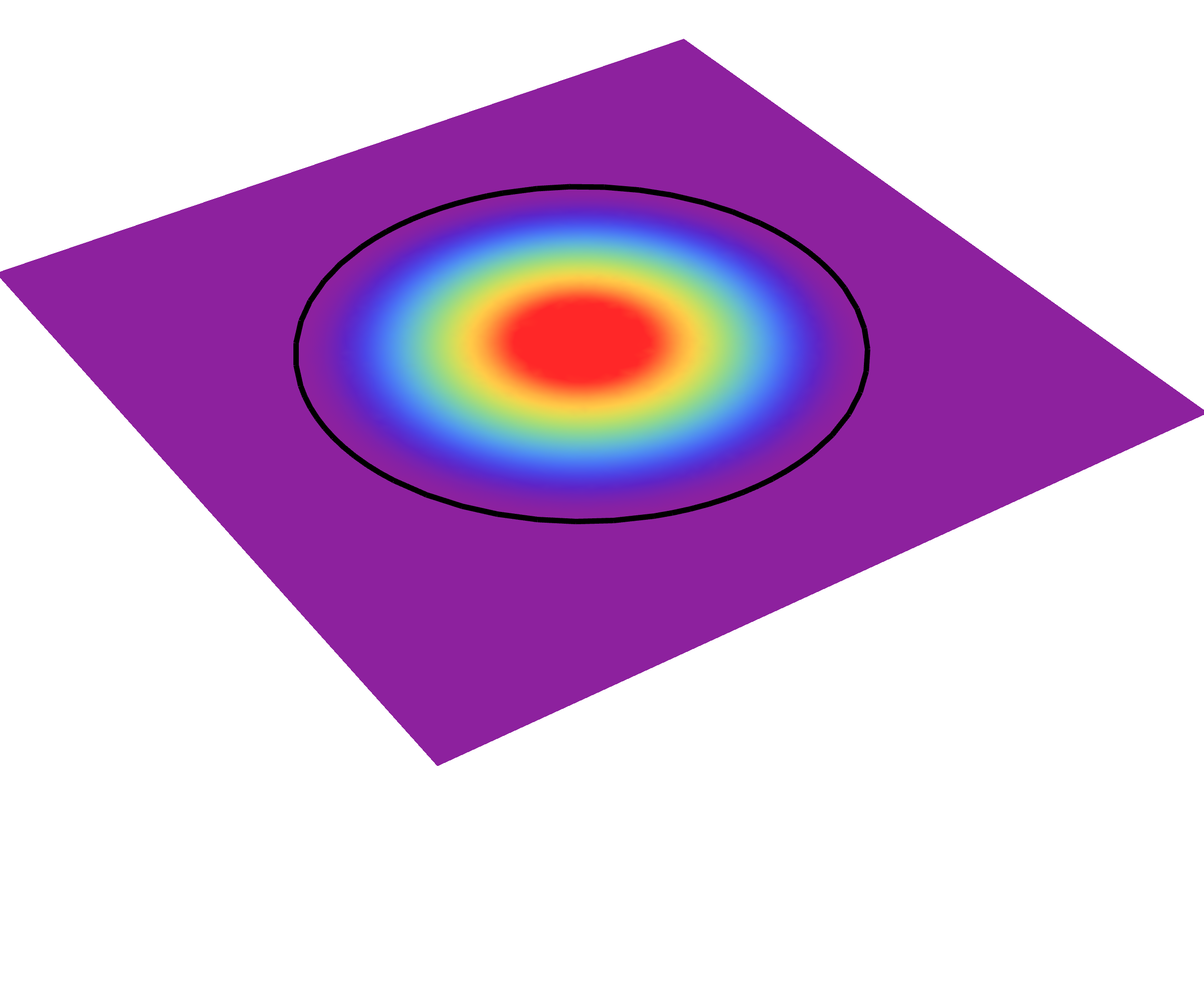}
\end{minipage}
\begin{minipage}{.48\textwidth}
    \centering
    \includegraphics[width=\linewidth]{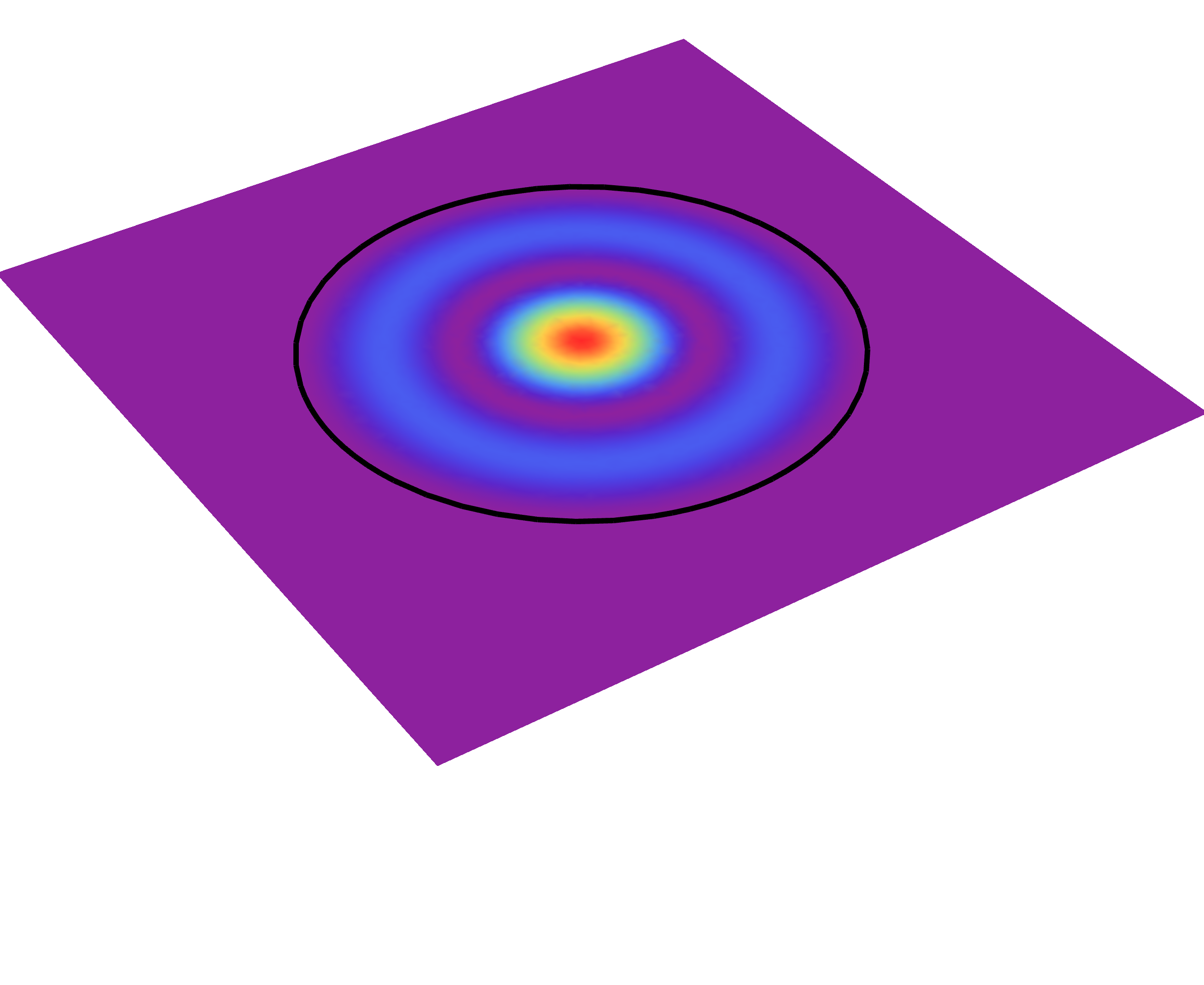}
\end{minipage}
\captionof{figure}{Schematic visualisation of the squared absolute value of two wave functions (eigenfunctions of the Laplacian) colour coded from violet (vanishing) to red on a flat two-dimensional background, here as plane embedded in three-dimensional space, and confined to a disk whose boundary is displayed in black.}
    \label{fig:WaveOnFlatBackground}
\end{figure}

In this section we take a more operational route towards constructing an instance of position uncertainty. Restricting the support of allowed wave functions in the Hilbert space underlying the quantum theory to a compact domain $\mathcal{D}$, \ie choosing it to be $\hil =L^2(\mathcal{D},\D^3x),$ we clearly localize them within a controllably sized setting. This can be achieved by imposing Dirichlet boundary conditions. Accordingly, all $\psi\in\hil$ have to satisfy $\psi|_{\partial \mathcal{D}}=0$ \ie vanish at the boundary and outside of it (see Fig. \ref{fig:WaveOnFlatBackground} for a visualisation). 

Any diffeomorphism invariant scale characterising the domain's extent would thus yield a measure of the position uncertainty. For example, we might use a function of the domain's volume
\begin{equation}
    V=\int_\mathcal{D} \D^d x.\label{volume}
\end{equation}
In particular, a position uncertainty of the dimensions of length is required, denoted by the symbol $\rho.$ Following this reasoning, we may choose
\begin{equation}
    \rho \propto\sqrt[d]{V}.\label{genposunc}
\end{equation}
In principle, this approach can be applied to any kind of domain. For reasons of simplicity, however, we will choose to work with geodesic balls and measure the position uncertainty through their radius. This information suffices to specify the Hilbert space which is about to be explored. Thus, we are all set to pose the problem whose solution yields the uncertainty relation.

\subsection{Eigenvalue problem and resulting inequality}\label{subsec:evprob}

The investigated quantum theory is set within a compact domain on a flat background manifold. Therefore, it can be shown, that the Laplacian, basically representing the squared conjugate momentum operator, is Hermitian and possesses a discrete spectrum \cite{Hilbert12}. Thus, its eigenvectors $\psi$ furnish an orthonormal basis of the Hilbert space $\hil.$ Evidently, they have to be solutions to the eigenvalue problem
\begin{subequations}
\label{evp}
\begin{align}
\Delta \psi + \lambda \psi	&=0~\text{within~}D\label{evp1}\\
\psi									&=0~\text{on~}\partial D.\label{evp2}
\end{align}
\end{subequations}
In $d$ dimensions, these eigenstates are characterized by $d$ quantum numbers, represented by the sole symbol $n$ to avoid index cluttering. An example of how the absolute values of the eigenstates may be distributed within a two-dimensional disk is given in Fig. \ref{fig:WaveOnFlatBackground}. A general state $\Psi$ can then be expressed as a linear combination of the eigenstates
\begin{equation}
    \Psi=\sum_{n}a_{n}\psi_{n}\label{genstaten}
\end{equation}
with the coefficients $a_{n}$ satisfying 
\begin{equation}
    \sum_{n}|a_{n}|^2=1.\label{coeflincomb}
\end{equation}
The standard deviation \eqref{flatstanddev} possesses a global minimum for a special state $\Psi_0$ such that 
\begin{equation}
    \sigma_p\left(\Psi\right)\geq\sigma_p\left(\Psi_0\right)\equiv\bar{\sigma}_p(\rho)\geq 0\label{uncgen}
\end{equation}
where $\rho,$ recall, denotes the measure of position uncertainty. Simple multiplication by $\rho$ yields the uncertainty relation in the usual form (as product of uncertainties)
\begin{equation}
    \sigma_p\rho\geq\bar{\sigma}_p(\rho)\rho.
\end{equation}
Thus finding the state $\Psi_0$ is all that is required to obtain the desired result. In general, it is hard to achieve that in a domain-independent fashion. Therefore, the domain is mainly chosen to be a geodesic ball below. First, however, we will give a general argument showing how the uncertainty relation scales in flat space. 

\subsection{Domain independent result in flat space}

Euclidean space has a simplifying advantage over all other Riemannian manifolds: Any kind of domain can be scaled up without ambiguities. Therefore, we are able to obtain a result for general domains following a simple conformal reasoning. Assume the measure of position uncertainty to be of the form \eqref{genposunc}. Increasing the volume of a general extended object in flat space by a constant factor $a^d,$ \ie transforming $\rho\rightarrow \tilde{\rho}=a\rho,$ is then equivalent to a conformal transformation of the metric $\delta_{ij}\rightarrow a^2\delta_{ij}.$ Correspondingly, the Laplacian transforms as $\Delta\rightarrow\Delta/a^2$ and therefore the $\text{n}^{\rm{th}}$ eigenvalue of the transformed Laplacian, denoted $\tilde{\lambda}_n$, satisfies
\begin{equation}
    \left(\frac{\Delta}{a^2}+\tilde{\lambda}_n\right)\tilde{\psi}_n=0.
\end{equation}
Evidently, the eigenvalues of the Laplacian transform accordingly: $\tilde{\lambda}_n=\lambda_n/a^2.$ Hence, we immediately see that
\begin{equation}
    \frac{\lambda_n}{\tilde{\lambda}_n}=\left(\frac{\tilde{\rho}}{\rho}\right)^2.
\end{equation}
As $C_n=\tilde{\lambda}_n\tilde{\rho}^2$ is just a dimensionless parameter independent of the scale $a,$ the entire dependence of the eigenvalues of the Laplacian on it has to be summarized in $\rho^{-2}.$ Thus, we obtain
\begin{equation}
    \lambda_n (a)=\frac{C_n}{\rho^2(a)}
\end{equation}
where the exact value of $C_n$ depends on the shape of the domain and the exact form of the position uncertainty $\rho$.

In general, the real and the imaginary parts of the eigenvalue problem \eqref{evp1} are colinear. This implies that the phase of its solutions $\psi_n$ can be removed by rotating the coordinate system. As the locally Euclidean frame is invariant under rotations, we can calculate the expectation value of the momentum operator in any of those related by a rotation. Thus, we can take the eigenfunctions of the Laplacian to be real.  However, the expectation value of the momentum operator with respect to any real wave function $\psi: {\rm I\!R}^3\rightarrow {\rm I\!R}$ vanishes as can be readily verified by
\begin{equation}
    \Braket*{\psi}{\hat{\pi}_a\psi}=\int\D\mu \psi\hat{\pi}_a\psi=-\int\D\mu \hat{\pi}_a(\psi)\psi=-\Braket*{\psi}{\hat{\pi}_a\psi}=0\label{vanmomrealwave}
\end{equation}
where $\D\mu$ stands for the integration measure in flat space ($\D\mu=\D^3x$ in Cartesian coordinates) and we used the Hermiticity of $\hat{\pi}_a$ as well as the boundary condition \eqref{evp2}. In fact, this statement continues to be true on curved backgrounds.

If the state saturating the uncertainty relation is an eigenvector of the Laplacian (represented as $n=1$), as was shown explicitly below for the geodesic ball in Ref. \cite{Petruzziello21}, the uncertainty relation in flat space thus reads
\begin{equation}
    \sigma_p\rho\geq\hbar C_1
\end{equation}
which shows the same scaling as Heisenberg's celebrated inequality. The value of $C_1$ is determined in the subsequent section for the specific choice of a geodesic ball as domain.

\subsection{Geodesic ball in three dimensions}\label{subsec:flat}

A geodesic ball $B_\rho$ is defined such that the geodesic distance $\sigma=\int\D s$ from its center $p_0$ to its boundary equals the radius $\rho.$ In flat space, of course, this is just the familiar ball. Rewritten in terms of spherical coordinates $\sigma^i=(\sigma,\chi,\gamma)$ and the explicit quantum numbers in three dimensions $n,l$ and $m,$ the eigenvalue problem \eqref{evp} becomes then
\begin{align}
   \left[ \partial_{\sigma}^2+\frac{2}{\sigma}\partial_{\sigma}+\frac{1}{\sigma^2}\left(\partial_{\chi}^2+\cot\chi\partial{\chi}+\sec^2\chi\partial_{\gamma}^2\right)+\lambda^{(0)}_{nlm}\right]\psi^{(0)}_{nlm}=&0\\
   \left.\psi^{(0)}_{nlm}\right|_{\sigma =\rho}=&0
\end{align}
where the superscript $(0)$ stands for the zeroth order of the perturbative expansion we perform below. This problem can be solved by seperation of variables yielding the result
\begin{align}
\psi^{(0)}_{nlm}	&=\sqrt{\frac{2}{\rho^3j^2_{l+1}(j_{l,n})}}j_{l}\left(j_{l,n}\frac{\sigma}{\rho}\right)Y^l_m(\chi,\gamma)\label{flatsol}\\
\lambda^{(0)}_{nlm}		&=\left(\frac{j_{l,n}}{\rho}\right)^2\label{flateig}
\end{align}
with the spherical harmonics $Y_m^l,$ the spherical Bessel function of first kind $j_l(x)$ and the $\text{n}^{\text{th}}$ zero of the spherical Bessel function of first kind $j_{l,n}.$ In particular, as shown in Ref. \cite{Petruzziello21}, the state saturating the uncertainty relation is the ground state of the Laplacian which reads
\begin{equation}
    \psi^{(0)}_{100}=\frac{1}{\sqrt{2\pi\rho}}\frac{\sin\left(\pi\frac{\sigma}{\rho}\right)}{\sigma}\label{unperteig100}
\end{equation}
and has eigenvalue $\lambda_{100}^{(0)}=\pi^2/\rho^2$ - a rather intuitive result because, being the ground state, it is the only distinguished state in the system. Therefore, we conclude that the ground state $\psi^{(0)}_{100}$ is indeed the state of lowest uncertainty in flat space
\begin{equation}
    \Psi_0=\psi_{100}^{(0)}.\label{lowestunc}
\end{equation}
We can infer from this result, that $C_1=\pi$ for geodesic balls yielding the flat-space uncertainty relation
\begin{equation}
    \sigma_p\rho\geq\pi\hbar.\label{flatres}
\end{equation}
This resembles but does not equal the inequality derived from the Robertson relation because those two describe different setups, which are non-linearly related. 

The power of the formalism introduced here, in the flat case akin to using a sledge-hammer to crack a nut, is shown to unfold at full strength below, where we obtain relativistic curvature induced corrections to the relation \eqref{flatres} perturbatively. 

\subsection{Generalization to curved space}

\begin{figure}[!htb]
\centering
\includegraphics[width=\linewidth]{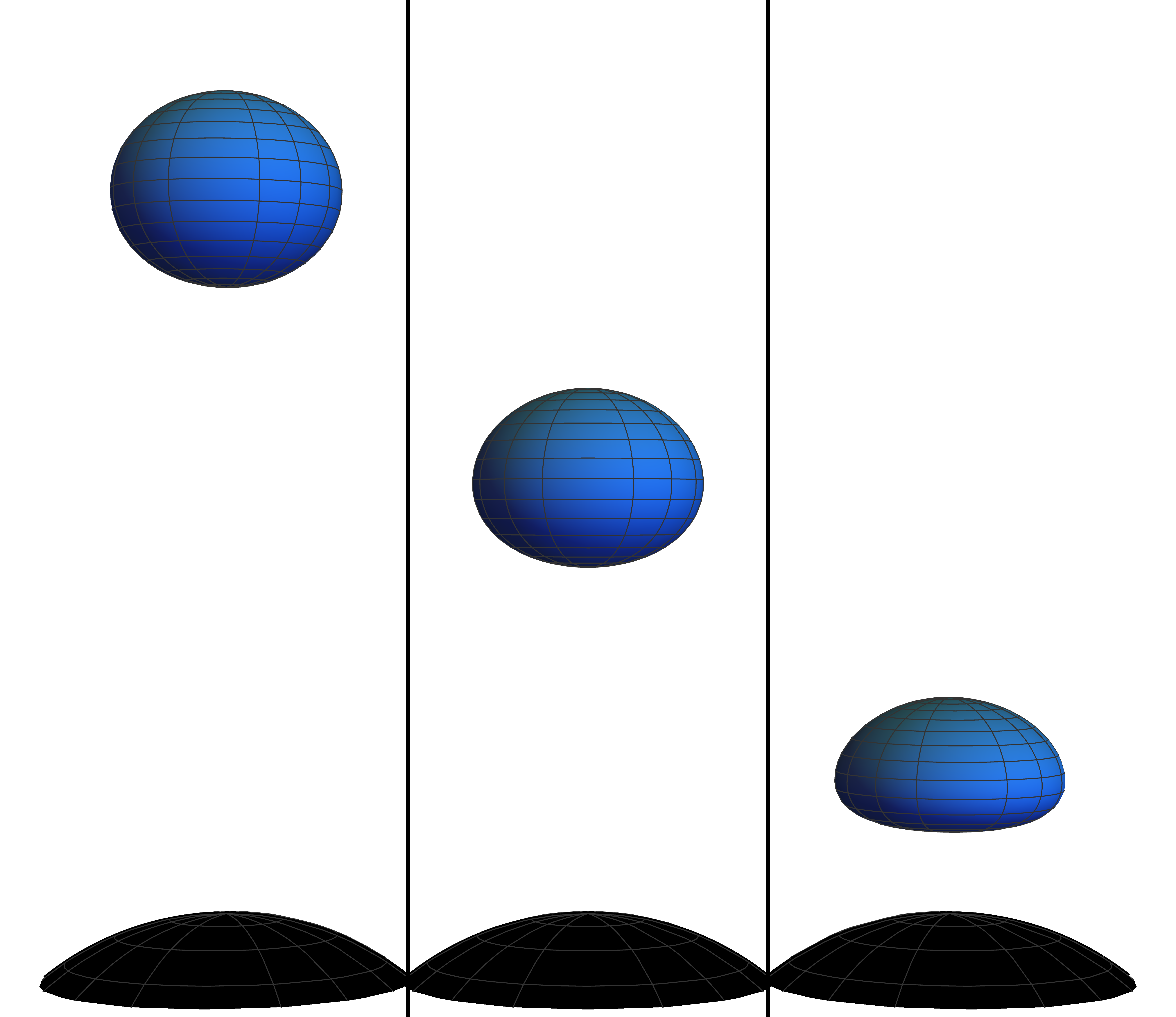}
\caption{Three geodesic balls whith equal geodesic radius $\rho=.4R_S$ but different distances from the center of symmetry $r_0=3.5R_S$ (left), $r_0=2.5R_S$ (mid) and $r_0=1.5R_S$ (right) on a spatial section of the Schwarzschild static patch characterized by the Schwarzschild radius $R_S$ and described in terms of Schwarzschild coordinates. Surfaces of geodesic balls are coloured blue while black hole horizons are indicated in black. \label{fig:balls}}
\end{figure}

\begin{figure}
\begin{minipage}{.49\textwidth}
    \centering
    \includegraphics[width=\linewidth]{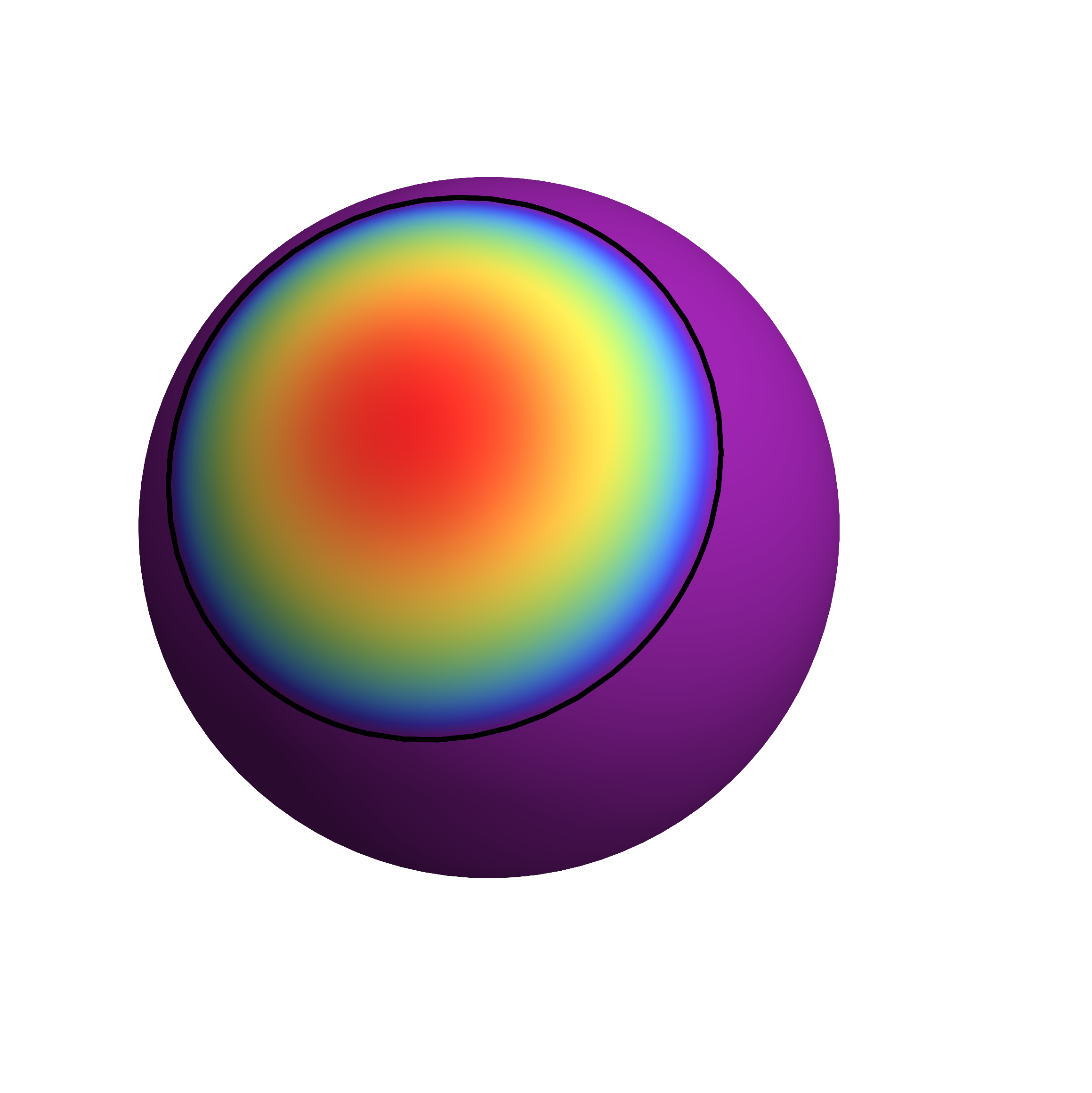}
\end{minipage}
\begin{minipage}{.49\textwidth}
    \centering
    \includegraphics[width=\linewidth]{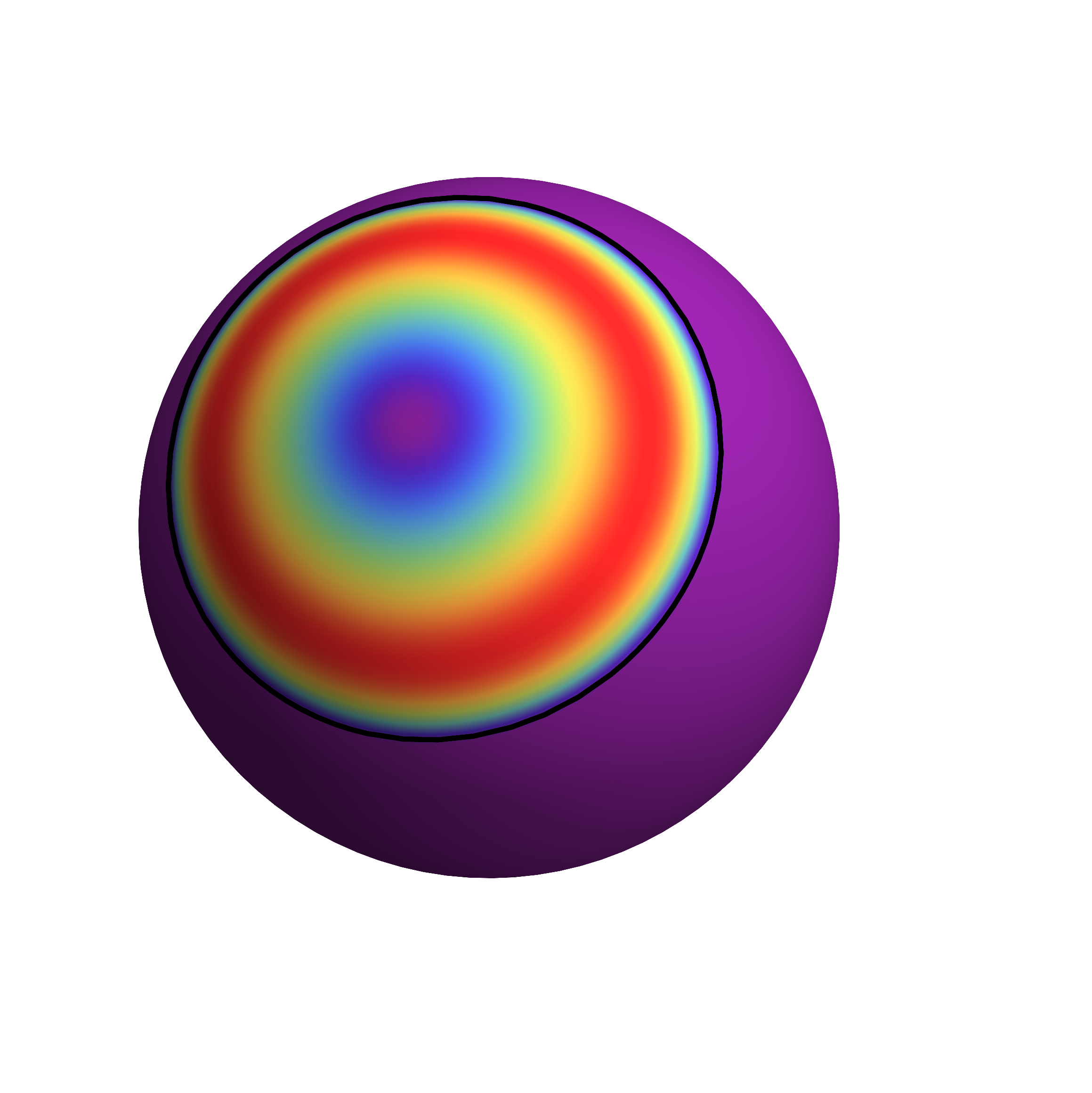}
\end{minipage}
\captionof{figure}{Schematic visualisation of the squared absolute value of two wave functions (eigenfunctions of the Laplacian) colour coded from violet (vanishing) to red on a sphere, imbedded in three-dimensional Euclidean space, and confined to a disk whose boundary is indicated in black.}
    \label{fig:WaveOnCurvedBackground}
\end{figure}

Almost all assertions made in this section swiftly generalize to nontrivial three-dimensional backgrounds. Geodesic balls, for example, continue to be well-defined objects with diffeomorphism invariant radius. However, they may not look like simple balls depending on the observer in question. An example of this variation is displayed in Fig. \ref{fig:balls} showing the distortion of a geodesic ball as it approaches a Schwarzschild horizon as seen from the static observer at spacelike infinity.
Furthermore, the eigenvalue problem \eqref{evp} basically stays the same. However, given a nontrivial background, the Laplacian has to be replaced by its covariant version, the Laplace-Beltrami operator. Thus, the setting is of the form displayed in Fig. \ref{fig:WaveOnCurvedBackground} for a two-dimensional spherical background.

However, we do not only live on a three-dimensional Riemannian manifold but in curved four-dimensional spacetime. In this context, there are a couple of subtleties in defining the physical momentum operator, \ie the momentum as it may be measured in experiments. These are dealt with in the following section.

\section{General relativistic momentum operator for massive particles}\label{sec:momop}

This section is devoted to the derivation of the operator corresponding to the momentum of a massive particle moving in a possibly curved spacetime. To this aim, the underlying classical quantity is treated first to subsequently provide its quantum mechanical counterpart.

\subsection{Classical physical momentum}

An approach, similar to the present one, was taken in Ref. \cite{Petruzziello21}, where the reader can find more details. Accordingly, the background metric $g_{\mu\nu}(x)$ may be parameterized as 
\begin{equation}
    \D s^2=-N^2\left(\D x^0\right)^2+h_{ij}(N^i\D x^0+\D x^i)(N^j\D x^0+\D x^j)
\end{equation}
in accordance with the ADM formalism \cite{Arnowitt60,Arnowitt08}, where the lapse function, the shift vector on the $d$-dimensional spacelike hypersurfaces of constant coordinate time $x^0$ are denoted as $N(x),$ $N^i(x)$ and $h_{ij}(x),$ respectively. Breaking the time reparametrization invariance of the action describing a particle on this background by choosing the affine parameter $\tau=x^0,$ the Lagrangian can then be expressed as
\begin{equation}
     L=-mN\sqrt{1-\beta^2}\label{LR1}
\end{equation}
with the analog of the squared ratio of velocity and speed of light in the curved-spacetime setting
\begin{equation}
    \beta^2\equiv \frac{\left(N^i+\dot{x}^i\right)\left(N^j+\dot{x}^j\right)h_{ij}}{N^2}.
\end{equation}
This quantity was expanded in in Ref. \cite{Petruzziello21} to obtain the nonrelativistic limit. The present treatment diverges from said approach at this point. Instead, the canonical momenta can directly be derived from the Lagrangian \eqref{LR1} yielding
\begin{equation}
    \pi_i\equiv \frac{\partial L}{\partial \dot{x}^i}=m\frac{G_{ij}(\dot{x}^j+N^i)}{\sqrt{1-\beta^2}}.\label{relmom1}
\end{equation}
This relation can be inverted to express the velocities in terms of the canonical momenta as
\begin{equation}
    \dot{x}^i=\frac{\sqrt{1-\beta^2}}{m}G^{ij}\pi_j-N^i
\end{equation}
with the effective spatial metric the particle is experiencing $G_{ij}=h_{ij}/N,$ which, importantly, does not necessarily equal the induced metric on spacelike hypersurfaces. This behaviour was already observed in the nonrelativistic case \cite{Petruzziello21}. There, the physical momentum is defined as 
\begin{equation}
    \left.p_i\right|_{\beta\ll 1}\equiv \left.mG_{ij}\dot{x}^j\right|_{\beta\ll 1}=\pi_i-mG_{ij}N^i \label{nonrelmom}
\end{equation}
to make it gauge invariant. In the relativistic context, though, we have to define the physical momentum differently multiplying the equivalent of the $\gamma$-factor to account for standard relativistic effects
\begin{equation}
    p_i\equiv\frac{mG_{ij}\dot{x}^j}{\sqrt{1-\beta^2}}=\pi_i-\sqrt{1+\frac{\pi^2}{Nm^2}}mG_{ij}N^j\label{relclassmom}
\end{equation}
with the squared canonical momentum $\pi^2=G^{ij}\pi_i\pi_j.$ For small canonical momenta with respect to the particle's mass, this clearly recovers eq. \eqref{nonrelmom}. In the ultrarelativistic limit, on the other hand it results in
\begin{equation}
    \left.p_i\right|_{\beta\gg 1}\equiv \left.mG_{ij}\dot{x}^j\right|_{\beta\ll 1}=\pi_i-\sqrt{\pi^2}G_{ij}N^i/\sqrt{N}, \label{ultrelmom}
\end{equation}
which, being independent of the mass, also applies to massless particles.

Importantly, the physical momentum \eqref{relclassmom} only equals the canonical momentum if the shift vector vanishes. A momentum measurement, thus, does not concern the canonical quantity in and of itself but the combination \eqref{relclassmom}. Therefore, uncertainty relations should in fact be obtained with respect to the physical momentum $p_i.$

\subsection{Quantum mechanical canonical momentum}

On a general Riemannian background, the position space representation with manifest diffeomorphism invariance necessitates the nontrivial integral measure
\begin{equation}
    \D\mu =\sqrt{G}\D^dx
\end{equation}
with the determinant of the effective metric $G.$ Assuming, that positions and canonical momenta satisfy the Heisenberg algebra \eqref{Heisalg} and that the operator representing the latter be Hermitian, it has to act on position space wave functions as \cite{DeWitt52}
\begin{equation}
    \hat{\pi}_i\psi=-i\hbar\left(\partial_i+\frac{1}{2}\Gamma^i_{ij}\right)\psi\equiv -i\hbar\nabla_i\psi
\end{equation}
where the last equality defines the covariant derivative acting on scalar densities of weight $1/2.$ Writing it this way, we concealed a subtlety, though.

The familiar treatment of vector operators in textbook quantum mechanics is not immediately generalizable to curved space. In particular, the expectation value of a vector operator, being an integral over a vector, is mathematically not well-defined. For example, we could describe the momentum in two distinct coordinate systems $x^i$ and $y^a$ expressing the components of a general one-form $\omega_i$ of the former in terms of the latter as
\begin{equation}
    \omega_i=\frac{\partial y^j}{\partial x^i}\omega_j. 
\end{equation}
Then, the expectation value of the canonical momentum operator in the coordinate system $x^i$ with respect to a general state $\ket{\psi}$ would read 
\begin{equation}
    \braket*{\hat{p}_i}=\braket*{\frac{\partial y^j}{\partial x^i}\hat{p}_j}=\int\D^dx\sqrt{g}\psi^*\left(-i\hbar\frac{\partial y^j}{\partial x^i}\nabla_j\right)\psi\neq \frac{\partial y^j}{\partial x^i}\int \D^dx\sqrt{g}\psi^*(-i\hbar\nabla_i)\psi
\end{equation}
where the transformation matrix $\partial y^a/\partial x^i,$ being position dependent, cannot be taken out of the integral. Thus, it is not diffeomorphism invariant as reflected by the inequality
\begin{equation}
    \braket*{\hat{\pi}_i}\neq \frac{\partial y^j}{\partial x^i}\braket*{\hat{\pi}_j}.
\end{equation}
This problem can be circumvented with the help of geometric calculus \cite{Hestenes84}. Expressed in this language, one-forms are expanded in terms of basis vectors $\gamma(x)^i$ which satisfy the generalized Clifford algebra
\begin{equation}
    \{\gamma^i,\gamma^j\}=2g^{ij}\label{Clifford}
\end{equation}
where the curly brackets stand for the anticommutator. The basis vectors themselves can be made independent of the position using the tetrad formalism \cite{Einstein28}. Define the \emph{vielbein} $e^i_a$ and its inverse $e^a_i$ such that
\begin{equation}
    g_{ij}\equiv e_i^ae_j^b\delta_{ab}\hspace{1cm} g^{ij}\equiv e^i_ae^j_b\delta^{ab}.\label{vielbein}
\end{equation}
Then, according to eq. \eqref{Clifford}, one can choose a basis such that $\gamma^a=e^a_i\gamma^i\neq\gamma^a(x)$ and
\begin{equation}
    \{\gamma^a,\gamma^b\}=2\delta^{ab}.\label{normalClifford}
\end{equation}
Applying all of this machinery and using the familiar Dirac-notation, a one-form $\slashed{\omega}$ can be expressed as $\slashed{\omega}=\gamma^i(x)\omega_i=\gamma^ae_a^i\omega_i.$ Vectors may be treated analogously. Then, the symetric contraction of a operator-valued vector $\hatslashed{V}=\gamma_i \hat{V}^i$ and an operator-valued one-form $\hatslashed{\omega}$ reads
\begin{equation}
    \left\{\hatslashed{V},\hatslashed{\omega}\right\}=\left\{\hat{V}^i,\hat{\omega}_i\right\}
\end{equation}
as expected. In short, expectation values of vectors should always be evaluated in a local Euclidean frame. On a flat background, this reduces to quantizing in Cartesian coordinates. This peculiar fact had already been stressed by Dirac in 1930 \cite{Dirac30}.

Define, thus, the conjugate momentum operator through its action on wave functions in position space \cite{Pavsic01}
\begin{equation}
    \hatslashed{\pi}\psi\equiv\gamma^i\hat{\pi}_i\psi=-i\hbar\gamma^i (x)\nabla_i \psi\label{newmomop},
\end{equation}
Its expectation value reads
\begin{equation}
    \braket*{\hatslashed{\pi}}=\int\D^4x\sqrt{G}\psi^*\left(-i\hbar\gamma^ae_a^i\nabla_i\psi\right)=\gamma^a\int\D^4x\sqrt{G}\psi^*\left(-i\hbar e_a^i\nabla_i\psi\right)=\gamma^a\braket*{e^i_a\hat{\pi}_i}.
\end{equation}
Here we could take the the basis vector out of the integral because, as was alluded to above, it is independent of the positions. Thus, it suffices to add in the \emph{vielbein}, \ie describe the momentum in a local Euclidean frame, to turn the expectation value of the momentum operator into a well-defined object. In Ref. \cite{Pavsic01}, the operator $\hatslashed{\pi}$ is shown to generate translations. Furthermore, it is proven that its square is proportional to the Laplace-Beltrami operator
\begin{equation}
    \hatslashed{\pi}^2\psi=\hat{\pi}^2\psi=- \hbar^2\frac{1}{\sqrt{G}}\partial_i\left(\sqrt{G}\partial^i\psi\right)\label{squaredmom}
\end{equation}
thereby claiming the correct relation to the free-particle Hamiltonian. Thus, it fulfills all the requirements to yield a position representation of the canonical momentum operator in curved space. However, it only describes the canonical momentum. The definition of the physical momentum operator bears subtleties in and of itself.

\subsection{Operator ordering ambiguities}\label{subsec:opord}

Quantization, provided we understand it as such in the first place, is not an injective map. In fact, given any classical function there is an infinite number of possible quantum operators corresponding to it. Consider, for example, the squared position in one dimension, $x^2,$ which could be derived as the classical limit of any operator of the symmetric form 
\begin{equation}
    F(\hat{p})\hat{x}F^{-2}(\hat{p})\hat{x}F(\hat{p})=\hat{x}^2-\hbar^2\left[\left(\frac{F'}{F}\right)^2+\frac{F''}{F}\right]\label{opordex}
\end{equation}
for a general real function $F.$ In the end, it is up to experiment to decide on the correct definition even though there may be theoretical reasons to prefer one ordering over another. For instance, the Laplace-Beltrami operator, a quantization of the classical function $g^{ij}(x)p_ip_j,$ apart from being backed by experiment, has the advantage that it is invariant under spatial diffeomorphisms as expected from the squared magnitude of the physical momentum. In the more primitive case of eq. \eqref{opordex}, however, there is just no reason to expect anything else than $\hat{x}^2$ to be the quantisation of the squared position. If no other principles can be found to guide the choice of ordering, it is thus intuitive to refrain from adding more ingredients. Furthermore, as they are supposed to be observables, the resulting operators have to be symmetric.

In comparison to the non-relativistic version \eqref{nonrelmom}, the relativistic momentum \eqref{relclassmom} mixes positions and canonical momenta. Thus, it is not of the primitive form featured in eq. \eqref{opordex}. Not specifying the exact prescription, its quantum mechanical counterpart can be written as
\begin{equation}
     \slashed{\hat{p}}=\hatslashed{\pi}-m\left(\sqrt{1+\frac{\hat{\pi}^2}{\hat{N}m^2}}\hatslashed{N}\right)_\Ord\label{nonpertrelmomop}
\end{equation}
where the subscript $\Ord$ stands for any symmetric ordering without addition of extra operators. Then, the inequality which is at the heart of the present work has to be derived from the standard deviation of the physical momentum operator
\begin{equation}
    \sigma_\hatslashed{p}=\sqrt{\braket*{\hatslashed{p}^2}-\braket*{\hatslashed{p}}^2}=\sqrt{\braket*{\hat{p}^2}-\delta^{ab}\braket*{\hat{p}_a}\braket*{\hat{p}_b}},
\end{equation}
\ie in a local Euclidean frame. The rest of the treatment is analogous to the one introduced in section \ref{sec:uncrel}. 

For the purpose of calculating this quantity, it is only required to show, that any symmetric term possibly appearing in an ordering, obeying said principles, of a power of the squared conjugate momentum and a position coordinate satisfies
\begin{equation}
    \frac{1}{2}\left(\hat{\pi}^{2J}\hat{x}^a\hat{\pi}^{2(\mathcal{N}-J)}+\hat{\pi}^{2(\mathcal{N}-J)}\hat{x}^a\hat{\pi}^{2J}\right)=\frac{1}{2}\left\{\hat{x}^a,\hat{\pi}^{2\mathcal{N}}\right\}+\left[\hat{\pi}^{2(\mathcal{N}-J)},\left[\hat{x}^a,\hat{\pi}^{2J}\right]\right]=\frac{1}{2}\left\{\hat{x}^a,\hat{\pi}^{2\mathcal{N}}\right\}
\end{equation}
with $J\in (0,1,\dots,\mathcal{N})$ and that similarly, once two coordinates are included
\begin{align}
    \frac{1}{2}\left(\hat{\pi}^{2J}\hat{x}^a\hat{\pi}^{2K}\hat{x}^b\hat{\pi}^{2[\mathcal{N}-(J+K)]}+\hat{\pi}^{2[\mathcal{N}-(J+K)]}\hat{x}^b\hat{\pi}^{2K}\hat{x}^a\hat{\pi}^{2J}\right)=&\frac{1}{2}\left\{\hat{x}^a\hat{x}^b,\hat{\pi}^{2\mathcal{N}}\right\}+i\hbar\left\{J\left[\hat{\pi}^{2[\mathcal{N}-(J+K)]}\hat{x}^b,\hat{\pi}^a\hat{\pi}^{2(J+K-1)}\right]\right.\nonumber\\
    &\left.-\left(J+K\right)\left[\hat{x}^a,\hat{\pi}^b\hat{\pi}^{2(\mathcal{N}-1)}\right]\right\}\\
    =&\frac{1}{2}\left\{\hat{x}^a\hat{x}^b,\hat{\pi}^{2\mathcal{N}}\right\}+\hbar^2\left\{J\hat{\pi}^{2(\mathcal{N}-1)}\delta^{ab}\right.+2\left[(J+K)(\mathcal{N}-1)\right.\nonumber\\
    &\left.\left.-J(J+K-1)\right]\hat{\pi}^{2(\mathcal{N}-2)}\hat{\pi}^a\hat{\pi}^b\right\}
\end{align}
with $J,K\in (0,1,\dots,\mathcal{N})$ and $J+K\in (0,1,\dots,\mathcal{N}).$ This implies that every function $f$, which is analytic on $\reals^+$ and therefore can be expanded nonsingularly for all elements in the spectrum of $\hat{\pi}^2,$ will satisfy
\begin{align}
    \left[f(\hat{\pi}^2)x^a\right]_\Ord=&\frac{1}{2}\left\{\hat{x}^a,f\left(\hat{\pi}\right)\right\}\label{ordering1}\\
        \left[f(\hat{\pi}^2)\hat{x}^a\hat{x}^b\right]_\Ord=&\frac{1}{2}\left\{\hat{x}^a\hat{x}^b,f\left(\hat{\pi}\right)\right\}+\hbar^2\left[\mathcal{G}\left(\hat{\pi}^2\right)\delta^{ab}+\tilde{\mathcal{G}}\left(\hat{\pi}^2\right)\hat{\pi}^a\hat{\pi}^b\right]\label{ordering2}
\end{align}
where the subscript $\Ord$ symbolizes a general symmetric ordering without adding extra operators and we introduced the two additional, not specified, but equally analytic functions $\mathcal{G}$ and $\tilde{\mathcal{G}}.$ Similar results hold for symmetric orderings of the forms $[f(\hat{\pi}^2)\{x^a\pi_b\}g]_\Ord$ and $[f(\hat{\pi}^2)\{x^ax^b\pi_c\pi_d\}g]_\Ord,$ where the curly brackets indicate, that the ordering in their interior is fixed. These identities suffice to show, that the resulting uncertainty relation is independent of operator ordering.

As position and momentum dependent operators appear within one square root in the expression \eqref{nonpertrelmomop}, the ordering has to be enforced at the perturbative level which, fortunately, is exactly what is required for the purpose of this paper. 

\subsection{Riemann normal coordinates}\label{subsec:RNC}

Assuming small position uncertainties, the geometry of the relevant neighbourhood of the underlying three-dimensional manifold may be approximated by describing the effective spatial metric in terms of Riemann normal coordinates $x^a$ defined around the point $p_0$ as
\begin{equation}
    G_{ab}\simeq \delta_{ab}-\frac{1}{3}R_{acbd}|_{p_0}x^cx^d
\end{equation} 
with the Riemann tensor $R_{acbd}.$ Furthermore, the lapse function and the shift vector may be expanded as
\begin{align}
    N\simeq&N|_{p_0}+\nabla_bN|_{p_0}x^b+\nabla_b\nabla_cN|_{p_0}x^bx^c\label{lapseexp}\\
    N^a\simeq&N^a|_{p_0}+\nabla_bN^a|_{p_0}x^b+\nabla_b\nabla_cN^a|_{p_0}x^bx^c.\label{shiftexp}
\end{align}
Both are constant at lowest order. Being a quantity derived from the metric, the canonical momentum operator expands as $\hatslashed{\pi}\simeq\hatslashed{\pi}_{(0)}+\hatslashed{\pi}_{(2)}.$ This implies that, applying eqs. \eqref{ordering1} and \eqref{ordering2} and the relation $[\hatslashed{\pi},\hat{x}^a]=[\hatslashed{\pi}_{(0)},\hat{x}^a],$ the physical momentum operator satisfies order by order
\begin{subequations}\label{physmomopexp}
\begin{align}
    \slashed{\hat{p}}_{(0)}=&\hatslashed{\pi}_{(0)}-m\slashed{N}|_{p_0}\sqrt{1+\hat{\Pi}^2}\label{p0}\\
    \slashed{\hat{p}}_{(1)}=&\frac{m}{2}\left(\frac{1}{2}\left.\slashed{N}\nabla_a\ln{N}\right|_{p_0}\left\{\hat{x}^a,\frac{\hat{\Pi}^2}{\sqrt{1+\hat{\Pi}^2}}\right\}-\nabla_a\slashed{N}|_{p_0}\left\{\hat{x}^a,\sqrt{1+\hat{\Pi}^2}\right\}\right)\label{p1}\\
    \slashed{\hat{p}}_{(2)}=&\hatslashed{\pi}_{(2)}+\frac{m}{4}\left[\left.\left(\nabla_a\slashed{N}\nabla_{b}\ln{N}+\slashed{N}\frac{\nabla_a\nabla_{b}N}{2N}\right)\right|_{p_0}\left\{\hat{x}^a\hat{x}^b,\frac{\hat{\Pi}^2}{\sqrt{1+\hat{\Pi}^2}}\right\}-\left.\slashed{N}\frac{\nabla_{a}N\nabla_{b}N}{4N^2}\right|_{p_0}\left\{\hat{x}^a\hat{x}^b,\frac{\hat{\Pi}^2\left(4+3\hat{\Pi}^2\right)}{\left(1+\hat{\Pi}^2\right)^{3/2}}\right\}\right.\nonumber\\
    &-\nabla_a\nabla_b\slashed{N}|_{p_0}\left\{\hat{x}^a\hat{x}^b,\sqrt{1+\hat{\Pi}^2}\right\}-\left.\frac{\slashed{N}}{2Nm^2}\right|_{p_0}\left\{\hat{\pi}^2_{(2)},\frac{1}{\sqrt{1+\hat{\Pi}^2}}\right\}\Bigg]+\slashed{\mathcal{G}}_{ab}\left(\hat{\pi}^2_{(0)}\right)\delta^{ab}+\tilde{\slashed{\mathcal{G}}}_{ab}\left(\hat{\pi}^2_{(0)}\right)\hat{\pi}_{(0)}^a\hat{\pi}_{(0)}^b\label{p2}
\end{align}
\end{subequations}
where we introduced the ordering-dependent tensor- and vector-valued functions $\slashed{\mathcal{G}}_{ab}$ and $\tilde{\slashed{\mathcal{G}}}_{ab},$ which are analytic on $\reals^+,$ and the operator $\hat{\Pi}^2\equiv \hat{\pi}_{(0)}^2/N|_{p_0}m^2.$ Its expectation value
\begin{equation}
    \braket*{\hat{\Pi}^2}=\frac{\beta^2_{(0)}}{1-\beta^2_{(0)}}
\end{equation}
measures the relativisticness of the given state at lowest order. Functions of $\hat{\Pi}^2$ can be expanded in the eigenstates of the Laplacian
\begin{equation}
    f\left(\hat{\Pi}^2\right)\equiv\sum_{n,l,m}f\left(\frac{\hbar^2\lambda^{(0)}_{nl}}{m^2N|_{p_0}}\right)\ket*{\psi_{nlm}^{(0)}}\bra*{\psi_{nlm}^{(0)}},
\end{equation}
where the states are represented as in eq. \eqref{flatsol} and the eigenvalues are provided in eq. \eqref{flateig}.  

In the nonrelativistic limit ($\hat{\Pi}^2\rightarrow 0$) the expansion of the physical momentum operator \eqref{physmomopexp} clearly recovers the expressions provided in Ref. \cite{Petruzziello21} as expected. Having thus obtained an expansion of the momentum operator around a point on our background manifold, it is time to tackle the main goal of this paper.

\section{Explicit solution}\label{sec:explicitsol}

As the theoretical subtleties have been settled, we can now proceed to derive the uncertainty relation for a general curved background. The result is first obtained analytically for a flat background to be further generalized to small perturbations around it as indicated by the expansion in the preceding section.

\subsection{Flat space}

\begin{figure}
    \centering
    \includegraphics[width=\linewidth]{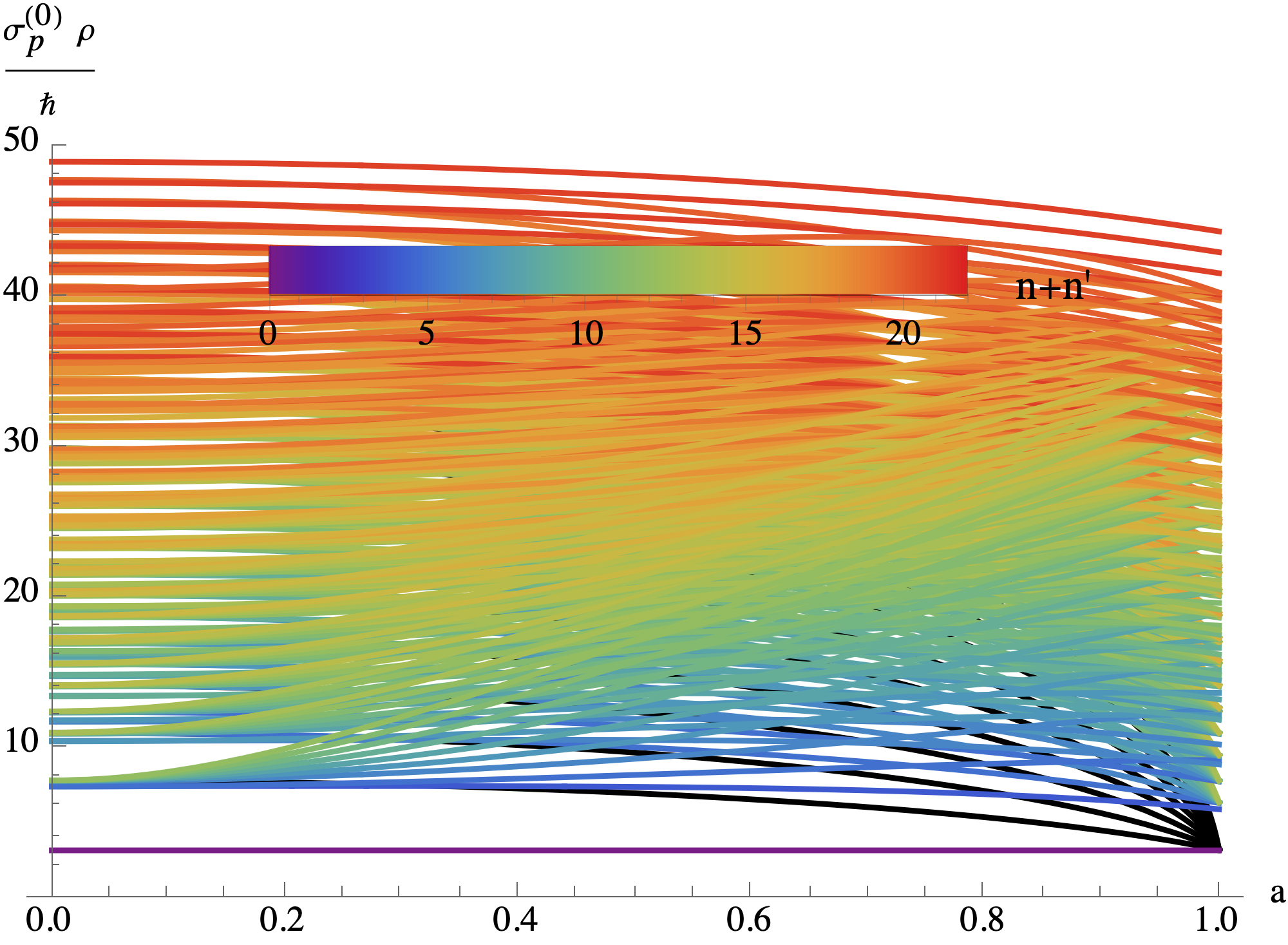}
    \caption{Lower bounds on the relativistic momentum uncertainties of all linear combinations of two eigenstates of the Laplacian \eqref{flatsol} with principal quantum numbers $(n,n')\leq (10,11)$ to lowest nonvanishing order as functions of the parameter $a\in[0,1]$ characterizing the relative weight and in units of $\hbar/\rho$. Black curves correspond to linear combinations including the ground state \eqref{unperteig100}, while the colour measures the sum of the principal quantum numbers $n$ and $n'$ for the others. The eigenvalue of the ground state is represented by the violet line.}
    \label{fig:smallestuncrel}
\end{figure}

As for non-relativistic particles, we begin with the uncertainty relation in flat space. In this case, the linear and squared momentum operators are given by eq. \eqref{p0} and as
\begin{equation}
    \hat{p}^2_{(0)}=\hat{\pi}^2_{(0)}-2m\slashed{N}|_{p_0}\left\{\hatslashed{\pi}_{(0)},\sqrt{1+\hat{\Pi}^2}\right\}+m^2N^aN_a|_{p_0}\left(1+\hat{\Pi}^2\right),
\end{equation}
Then, the variance of the momentum operator $\sigma_p^2$ can be expressed as
\begin{equation}
    \var{p}^{(0)}=\var{\pi}^{(0)}+\var{p}^{(0)}_{\text{rel}}\label{varsplitflat}
\end{equation}
where the global minimum of $\var{\pi}^{(0)},$ stemming from the ground state of the Laplacian $\psi_{100},$ was found in section \ref{subsec:flat} and we introduced the relativistic correction
\begin{equation}
    \var{p}^{(0)}_{\text{rel}}=2m\slashed{N}|_{p_0}\left(\braket*{\hatslashed{\pi}^{(0)}}\left\langle\sqrt{1+\hat{\Pi}^2}\right\rangle-\braket*{\hatslashed{\pi}^{(0)}\sqrt{1+\hat{\Pi}^2}}\right)+m^2\left.N^aN_a\right|_{p_0}\left[\braket*{1+\hat{\Pi}^2}-\braket*{\sqrt{1+\hat{\Pi}^2}}^2\right].
\end{equation}
Clearly, the first two terms can decrease the uncertainty when $\braket{\hatslashed{\pi}_{(0)}}\neq 0$ \ie for superpositions of eigenstates of the Laplacian with relative phase, the kind which was treated in Ref. \cite{Petruzziello21}. Expressed as a linear combination $\Psi$ (\cf eq. \eqref{genstaten}) of the eigenstates of the Laplacian and applying eqs. \eqref{evp1} and \eqref{vanmomrealwave}, it becomes
\begin{align}
    \var{p}^{(0)}_{\text{rel}}=&2m\slashed{N}|_{p_0}\sum_{n\neq  n'}\text{Re}\left(a^*_{n'}a_n\Braket*{\psi^{(0)}_{n'}}{\hatslashed{\pi}_{(0)}\psi^{(0)}_n}\right)\left(\sum_{n''} |a_{n''}|^2\sqrt{1+\frac{\lambdabar_C^2\lambda_{n''}}{N|_{p_0}}}- \sqrt{1+\frac{\lambdabar_C^2\lambda_n}{N|_{p_0}}}\right)\\
    \geq&-2m||\slashed{N}||_{p_0}\sum_{n\neq n'}|a_{n}||a_{n'}|\left|\left|\text{MaxRe}\left(e^{i\Delta\phi_{nl,n'l'}}\Braket*{\psi^{(0)}_{n'}}{\hatslashed{\pi}_{(0)}\psi^{(0)}_n}\right)\right|\right|\left|\sum_{n''} \left(|a_{n''}|^2-\delta_{n''n}\right)\sqrt{1+\frac{\lambdabar_C^2\lambda_{n''}}{N|_{p_0}}}\right|\nonumber\\
    &+m^2\left. N^aN_a\right|_{p_0}\left[1+\sum_n\left|a_n\right|^2\left(\frac{\lambdabar_C^2\lambda_n}{N|_{p_0}}-\sum_{n'}\left|a_{n'}\right|^2\sqrt{1+\frac{\lambdabar_C^2\lambda_{n}}{N|_{p_0}}}\sqrt{1+\frac{\lambdabar_C^2\lambda_{n'}}{N|_{p_0}}}\right)\right]
\end{align}
where $\lambdabar_C$ stands for the reduced Compton wave length and MaxRe indicates a choice of relative phase $\Delta\phi_{nln'l'}$ between the coefficients $a_n,$ $a_{n'}$ such that the real part of the resulting quantity is maximized. In the nonrelativistic limit the corrections multiply the factor $\sum_{n''}(|a_{n''}|^2-\delta_{n''n})=0$ and the contribution vanishes as expected. After some straight-forward yet tedious algebra displayed in appendix \ref{app_maxre} and recovering all quantum numbers in three dimensions, we can estimate
\begin{align}
    \var{p}^{(0)}_{\text{rel}}\geq& -\frac{2\hbar^2}{\rho\lambdabar_C}\sqrt{N^aN^bG_{ab}}\big|_{p_0}\sum_{n,l\neq n',l'}|a_{n}||a_{n'}|\frac{j_{l,n}j_{l',n'}}{|j^2_{l,n}-j^2_{l',n'}|}\left|\sum_{n'',l''} \left(|a_{n'',l''}|^2-\delta_{n''n}\delta_{l''l}\right)\sqrt{1+\frac{\lambdabar_C^2j^2_{l'',n''}}{\rho^2N|_{p_0}}}\right|\nonumber\\
    &+\left. N^aN_a\right|_{p_0}\left[1+\sum_n\left|a_n\right|^2\left(\frac{\lambdabar_C^2j_{l,n}^2}{\rho^2N|_{p_0}}-\sum_{n'}\left|a_{n'}\right|^2\sqrt{1+\frac{\lambdabar_C^2j_{l,n}^2}{\rho^2N|_{p_0}}}\sqrt{1+\frac{\lambdabar_C^2j_{l',n'}^2}{\rho^2 N|_{p_0}}}\right)\right].
\end{align}
As the non-relativistic case has been treated already in Ref. \cite{Petruzziello21}, a possible change in the state of smallest uncertainty should be expected to result in the ultra-relativistic limit \ie for states for which $\lambdabar_Cj_{l'',n''}/N|_{p_0}\rho^2\gg 1.$ Then the relativistic correction becomes approximately
\begin{align}
    \frac{\rho^2}{\hbar^2}\left.\var{p}^{(0)}_{\text{rel}}\right|_{\braket{\hat{\Pi}}\gg 1}\gtrsim& -2\sqrt{\frac{N_aN^a}{N}}\Bigg|_{p_0}\sum_{n,l\neq n',l'}|a_{n}||a_{n'}|\frac{j_{l,n}j_{l',n'}}{|j^2_{l,n}-j^2_{l',n'}|}\left|\sum_{n'',l''} \left(|a_{n'',l''}|^2-\delta_{nn''}\delta_{ll''}\right)j_{l'',n''}\right|\nonumber\\
    &+\left. \frac{N^aN_a}{N}\right|_{p_0}\sum_n\left|a_n\right|^2\left(j_{l,n}^2-\sum_{n'}\left|a_{n'}\right|^2j_{l,n}j_{l',n'}\right).
\end{align}
This implies, that the sum of both relevant contributions to the variance \eqref{varsplitflat} satisfies at the ultrarelativistic level
\begin{align}
    \frac{\rho^2}{\hbar^2}\left.\var{p}^{(0)}\right|_{\braket{\hat{\Pi}}\gg 1}\gtrsim& -2\sqrt{\frac{N_aN^a}{N}}\Bigg|_{p_0}\sum_{n,l\neq n',l'}|a_{n}||a_{n'}|\frac{j_{l,n}j_{l',n'}}{|j^2_{l,n}-j^2_{l',n'}|}\left|\sum_{n'',l''} \left(|a_{n'',l''}|^2-\delta_{nn''}\delta_{ll''}\right)j_{l'',n''}\right|\nonumber\\
    &+\sum_n\left|a_n\right|^2\left[j_{l,n}^2\left(1+\left. \frac{N^aN_a}{N}\right|_{p_0}\right)-\left. \frac{N^aN_a}{N}\right|_{p_0}\sum_{n'}\left|a_{n'}\right|^2j_{l,n}j_{l',n'}\right].
\end{align}
As the transition amplitude $\Braket{\psi_{nlm}}{\hatslashed{\pi}_{(0)}\psi_{n'l'm'}}$ is only nonvanishing if $\Delta l=|l'-l|=1,$ which is shown in appendix \ref{app_maxre}, the effect of linearly combining more than two eigenstates of the Laplacian cannot be stronger than just adding two of them. Thus, we can consider only the former without loss of generality. Then, we can define the relative weight $a\equiv|a_{n,l}|=\sqrt{1-|a_{n',l'}|^2}$ leading to the relation
\begin{align}
    \left.\var{p}^{(0)}\right|_{\braket{\hat{\Pi}}\gg 1}\geq&\frac{\hbar^2}{\rho^2}\left[ A(a)-B(a)\left.\sqrt{\frac{N^aN_a}{N}}\right|_{p_0}+C(a)\left.\frac{N^aN_a}{N}\right|_{p_0}\right]
\end{align}
where, denoting the quantum numbers of the two states as $n,l$ and $n',l'$ by a slight abuse of notation, we introduced the functions of the parameter
\begin{align}
    A(a)=&a^2j_{l,n}^2+(1-a^2)j_{l',n'}^2\\
    B(a)=&2a\sqrt{1-a^2}\left|1-2a^2\right|
    \frac{j_{l',n'}j_{l,n}}{|j^2_{l',n'}-j^2_{l,n}|}\left|j_{l,n}-j_{l',n'}\right|\\
    C(a)=&a^2(1-a^2)\left(j_{l,n}-j_{l',n'}\right)^2.
\end{align}
As a function of the shift vector and the lapse function the uncertainty clearly has a global minimum at $\sqrt{N^aN_a/N}|_{p_0}=\frac{B}{2C}.$ Thus, we can estimate
\begin{align}
    \left.\var{p}^{(0)}\right|_{\braket{\hat{\Pi}}\gg 1}\geq&\frac{\hbar^2}{\rho^2}\left[ A-\frac{B^2}{4C}\right]=\frac{\hbar^2}{\rho^2}\left[a^2j_{l,n}^2+(1-a^2)j_{l',n'}^2-\left(1-2a^2\right)^2\frac{j_{l',n'}j_{l,n}}{|j^2_{l',n'}-j^2_{l,n}|}\right].
\end{align}
The resulting uncertainties as functions of the parameter $a$ are plotted for all eigenfunctions of the Laplacian \eqref{flatsol} characterized by quantum numbers $n\leq 10,n'\leq 11$ in figure \ref{fig:smallestuncrel}.
Not a single one of those states has a smaller uncertainty than the ground state of the Laplacian $\psi_{100}^{(0)}$ defined in Eq. \eqref{unperteig100}. Those mixing with the ground state as 
\begin{equation}
    \Psi=ae^{i\Delta\phi}\psi_{100}^{(0)}+\sqrt{1-a^2}\psi_{nlm}^{(0)},
\end{equation}
coloured black, for example, only reach their minimum value at $a=1.$ Furthermore, the difference only grows with increasing $n+n'$ as can be inferred from the colour of the other graphs. 

To put it in a nutshell, the ground state of the Laplacian $\psi_{100}^{(0)}$ remains the state of smallest uncertainty, \ie we have found
\begin{equation}
    \Psi_0^{(0)}=\psi_{100}^{(0)}
\end{equation}
and we recover the inequality \eqref{flatres} in the relativistic setting. This result will be modified by gradually adding in curvature.

\subsection{Corrections}

As perturbative corrections are comparably small by definition, the fact, that the ground state of the Laplacian uniquely saturates the uncertainty relation, carries over to the slightly curved setting in general meaning, that
\begin{equation}
    \Psi_0=\psi_{100}.
\end{equation}
This implies, that eq. \eqref{vanmomrealwave} continues to hold to all orders in the expansion. Furthermore, the integration measure in flat space, with respect to which corrections to expectation values are computed (\cf Ref. \cite{Dabrowski20}), is even in the radial coordinate. Hence, all expectation values of operators, which are odd in the sum of the numbers of positions and unperturbed momenta vanish. Therefore, the derivation of the correction
\begin{equation}
    \var{p}^{(1)}(\psi_{100})=\Braket*{\psi_{100}^{(0)}}{\hat{p}^2_{(1)}\psi_{100}^{(0)}}_0-2\Braket*{\psi_{100}^{(0)}}{\hat{p}_{(1)}\psi_{100}^{(0)}}_0\Braket*{\psi_{100}^{(0)}}{\hat{p}_{(0)}\psi_{100}^{(0)}}_0,
\end{equation}
where the subscript $0$ at expectation values stands for the flat-space integration measure, simplifies significantly.

In fact, at this order all possibly arising corrections to $\braket{\hatslashed{p}}$ are even in the momenta and odd in the coordinates implying $\braket{\hatslashed{p}_{(1)}}=0.$ Most of the contributions to $\braket{\hat{p}^2}$ vanish for the same reason. The remaining terms assume the form
\begin{equation}
    \Braket*{\psi_{100}^{(0)}}{\hat{p}_{(1)}^2\psi_{100}^{(0)}}_0\propto \Braket*{\psi_{100}^{(0)}}{\left\{\hat{x}^a\hat{\pi}_b\right\}\psi_{100}^{(0)}}_0.
\end{equation}
This can be shown to equal zero applying the canonical commutation relations \eqref{Heisalg} and computing explicitly that
\begin{equation}
    \Braket*{\psi_{100}^{(0)}}{\hat{x}^a\hat{\pi}_b\psi_{100}^{(0)}}_0=\frac{i\hbar}{2}\delta^a_b.\label{xpiint}
\end{equation}
To put it in a nutshell, there are no first-order corrections to the uncertainty relation. In order to obtain curvature-induced contributions, it is necessary to treat the system at higher order.

The second-order contribution to the variance of the momentum operator
\begin{equation}
    \var{p}^{(2)}=\braket*{\left(\hatslashed{p}^{(1)}\right)^2}_0-\braket*{\hatslashed{p}^{(1)}}^2_0+\braket*{\left\{\hatslashed{p}^{(2)},\hatslashed{p}^{(0)}\right\}}_0-2\braket*{\hatslashed{p}^{(2)}}_0\braket*{\hatslashed{p}^{(0)}}_0,
\end{equation}
for example, yields meaningful terms.  Still, it simplifies considerably taking into account generic cancellations. The second term was shown to vanish when treating the calculations at first order. Furthermore, eq. \eqref{p0} implies that the third and the fourth terms largely cancel for all eigenstates of the Laplacian leaving us with
\begin{equation}
    \var{p}^{(2)}\left(\psi_{nlm}\right)=\braket*{\left(\hatslashed{p}^{(1)}\right)^2}_0+\braket*{\left\{\hatslashed{p}^{(2)},\hatslashed{\pi}^{(0)}\right\}}_0+\braket*{\left[\hatslashed{p}^{(0)}-\hatslashed{\pi}^{(0)},\hatslashed{p}^{(2)}\right]}_0.
\end{equation}
All contributions to $\hatslashed{p}^{(2)}$ as given in eq. \eqref{p2} except for $\hatslashed{\pi}^{(2)}$ are even in $\hat{\pi}_a$ and $\hat{x}^b$ while $\hatslashed{\pi}^{(0)}$ is evidently odd. Thus, when evaluated respective to the ground state of the Laplacian, this sum experiences a further simplification to read
\begin{equation}
    \var{p}^{(2)}\left(\psi_{100}\right)=\braket*{\hat{\pi}^2_{(2)}}_0+\braket*{\left(\hatslashed{p}^{(1)}\right)^2}_0+\braket*{\left[\hatslashed{p}^{(0)}-\hatslashed{\pi}^{(0)},\hatslashed{p}^{(2)}\right]}_0.\label{var2ndaftcanc}
\end{equation}
The first term appearing at the right-hand side just equals $\var{\pi}^{(2)}$ as derived in Ref. \cite{Dabrowski20} yielding
\begin{equation}
    \braket*{\pi_{(2)}^2}=-\frac{R|_{p_0}}{6}.
\end{equation}
with the Ricci scalar $R$ derived from the effective metric $G_{ab},$ while the correction obtained in Ref. \cite{Petruzziello21} is hidden in the second term. Making use of eq. \eqref{p1}, this expectation value is of the form
\begin{align}
    \braket*{\left(\hatslashed{p}^{(1)}\right)^2}_0=&\frac{m^2}{4}\braket*{\left\{\hat{x}^a,\slashed{F}_a\left(\hat{\Pi}^2\right)\right\}^2}_0\\
    =&m^2\left\{F_{ac}F_b^{~c}\left(\Pi^2\right)\braket*{x^ax^b}_0+\rho^2\frac{\Pi^2}{\pi^2}\left[\left(F_{ac}F^{ac}\right)'-\frac{\rho^2}{\hbar^2}\frac{\Pi^2}{\pi^2}\left(2F_{ac}F_{bc}''+F_{ac}'F_{bc}'\right)\braket*{\hat{\pi}^a\hat{\pi}^b}\right]\right\}\label{p1sqexprel}
\end{align}
where $\Pi^2=\hbar^2\pi^2/\rho^2m^2N|_{p_0}=\pi^2\lambdabar_C^2/\rho^2N|_{p_0},$ with the reduced Compton wavelength $\lambda_C\equiv \hbar/m,$ denotes the relativsticness of the ground state and we introduced the dimensionful, tensor-valued function
\begin{equation}
F_{ac}\left(\hat{\Pi}^2\right)=\frac{1}{2}N_c\nabla_a\ln{N}|_{p_0}\frac{\hat{\Pi}^2}{\sqrt{1+\hat{\Pi}^2}}-\nabla_aN_c|_{p_0}\sqrt{1+\hat{\Pi}^2}.
\end{equation}
The required expectation values can be evaluated explicitly yielding
\begin{align}
    \Braket*{\psi_{100}^{(0)}}{\hat{x}^a\hat{x}^b\psi_{100}^{(0)}}_0=&\frac{2\pi^2-3}{18\pi^2}\rho^2\delta^{ab}\equiv \frac{\xi}{2}\rho^2\delta^{ab}\label{doublexint}\\
    \Braket*{\psi_{100}^{(0)}}{\hat{\pi}^a\hat{\pi}^b\psi_{100}^{(0)}}_0=&\frac{\hbar^2\pi^2}{3\rho^2}\delta^{ab}\label{doublepiint}
\end{align}
where the third equality of the first equation defines the mathematical constant $\xi\sim\Ord{(10^{-1})}.$ Plugging these explicit results back in eq. \eqref{p1sqexprel}, we obtain
\begin{equation}
    \braket*{\left(\hatslashed{p}^{(1)}\right)^2}_0=m^2\rho^2\left\{\frac{\xi}{2} F_{ac}F^{ac}+\frac{\Pi^2}{\pi^2}\left[\frac{1}{2}\left(F_{ac}F^{ac}\right)'-\frac{\Pi^2}{3}\left(2F_{ac}''F^{ac}+F_{ac}'F^{\prime ac}\right)\right]\right\}.
\end{equation}
Fortunately, the third term of eq. \eqref{var2ndaftcanc} turns out to be such that the dependence on the ordering in the second-order correction to the momentum operator \eqref{p2} exactly cancels. Resultingly, this contribution can be expressed as
\begin{align}
    \braket*{\left[\hatslashed{p}^{(0)}-\hatslashed{\pi}^{(0)},\hatslashed{p}^{(2)}\right]}_0=&\Braket*{\psi_{100}^{(0)}}{\hat{\pi}^a\hat{\pi}^b\psi_{100}^{(0)}}_0\left[\frac{m^2\rho^4}{\hbar^2\pi^4}\frac{\Pi^4}{\sqrt{1+\Pi^2}}\left(\frac{\tilde{F}_{ab}\left(\Pi^2\right)}{1+\Pi^2}-\tilde{F}_{ab}'\left(\Pi^2\right)\right)+\frac{\rho^2}{6\pi^2}\frac{\Pi^2}{1+\Pi^2}\left.\frac{N^cN_cR_{ab}}{N}\right|_{p_0}\right]\\
    =&\frac{\rho^2m^2}{3\pi^4}\frac{\Pi^2}{\sqrt{1+\Pi^2}}\left(\frac{\tilde{F}_a^a}{1+\Pi^2}-\tilde{F}_a^{\prime a}\right)+\frac{\hbar^2}{36}\frac{\Pi^2}{1+\Pi^2}\left.\frac{N^cN_cR}{N}\right|_{p_0}
\end{align}
where we introduced the dimensionful, tensor-valued function
\begin{align}
    \tilde{F}_{ab}=&\left.\frac{ N_c}{2}\left[\frac{1}{2}\nabla_a\left(N^c\nabla_{b}\ln{N}\right)\frac{\hat{\Pi}^2}{\sqrt{1+\hat{\Pi}^2}}-\nabla_a\nabla_bN^c\sqrt{1+\hat{\Pi}^2}-\frac{1}{8}N^c\nabla_{a}\ln{N}\nabla_{b}\ln{N}\frac{\hat{\Pi}^2\left(4+3\hat{\Pi}^2\right)}{\left(1+\hat{\Pi}^2\right)^{3/2}}\right]\right|_{p_0}.
\end{align}
Plugging all those terms back into eq. \eqref{var2ndaftcanc}, the correction to the variance of the momentum operator reads
\begin{align}
    \var{p}^{(2)}=&\frac{\hbar^2}{2}\left[\left.\frac{ R}{9}\left(\frac{N_aN^a}{N}\frac{\Pi^2}{1+\Pi^2}-3\right)\right|_{p_0}+\mathcal{F}_1\left(\Pi^2\right)\left.\frac{N_bN^b}{N}\nabla_{a}\ln{N}\nabla^{a}\ln{N}\right|_{p_0}-\mathcal{F}_2\left(\Pi^2\right)\left.\frac{N^b\nabla_{a}N_b}{N}\nabla^{a}\ln{N}\right|_{p_0}\right.\nonumber\\
    &+\left.\mathcal{F}_3\left(\Pi^2\right)\left.\frac{\nabla_{a}N_b\nabla^aN^b}{N}\right|_{p_0}-\mathcal{F}_4\left(\Pi^2\right)\left.\frac{N^b\Delta N_b}{N}\right|_{p_0}+\mathcal{F}_5\left(\Pi^2\right)\left.\frac{N_bN^b}{N}\Delta\ln{N}\right|_{p_0}\right]
\end{align}
where we introduced the functions of relativisticness
\begin{align}
    \mathcal{F}_1=&\frac{\pi^2}{4}\xi\frac{\Pi^2}{1+\Pi^2}+\frac{\Pi^2\left(-4+5\Pi^2+2\Pi^4\right) }{12\left(1+\Pi^2\right)^3},\\
    \mathcal{F}_2=&\pi^2\xi+1-\frac{\Pi^2}{3\left(1+\Pi^2\right)^2},\\
    \mathcal{F}_3=&\pi^2\xi\frac{1+\Pi^2}{\Pi^2}+1+\frac{1}{6}\frac{\Pi^2}{1+\Pi^2},\\
    \mathcal{F}_4=&\frac{1}{6}\frac{\Pi^2}{1+\Pi^2},\\
    \mathcal{F}_5=&\frac{1}{12}\frac{\Pi^2\left(4+\Pi^2\right)}{\left(1+\Pi^2\right)^2}.
\end{align}
The final result of this paper thus reads
\begin{align}
    \sigma_p\rho\geq& \pi\hbar\left\{1+\frac{\rho^2}{4\pi^2}\left[\left.\frac{ R}{9}\left(\frac{N_aN^a}{N}\frac{\Pi^2}{1+\Pi^2}-3\right)\right|_{p_0}+\mathcal{F}_1\left(\Pi^2\right)\left.\nabla_{a}\ln{N}\nabla^{a}\ln{N}\frac{N_bN^b}{N}\right|_{p_0}-\mathcal{F}_2\left(\Pi^2\right)\left.\nabla^{a}\ln{N}\frac{\nabla_{a}N_bN^b}{N}\right|_{p_0}\right.\right.\nonumber\\
    &+\left.\left.\mathcal{F}_3\left(\Pi^2\right)\left.\frac{\nabla_{a}N_b\nabla^aN^b}{N}\right|_{p_0}-\mathcal{F}_4\left(\Pi^2\right)\left.\frac{N^b\Delta N_b}{N}\right|_{p_0}+\mathcal{F}_5\left(\Pi^2\right)\left.\frac{N_bN^b}{N}\Delta\ln{N}\right|_{p_0}\right]\right\}.
\end{align}
Undoubtedly, this is a quite involved expression. Therefore, it is instructive to consider its asymptotic behaviour. On the one hand, for $\Pi\ll 1,$ implying the non-relativistic limit,  we recover the relation derived in Ref. \cite{Petruzziello21}
\begin{equation}
    \sigma_p\rho\geq\pi\hbar\left[1-\rho^2\left.\left(\frac{R}{12\pi^2}-\xi\frac{\rho^2}{\lambda_C^2}\nabla_aN_b\nabla^aN^b\right)\right|_{p_0}\right]\label{nonrelunc}
\end{equation}
as expected. Ultra-relativistic particles satisfying $\Pi\gg 1$, on the other hand, obey the uncertainty relation
\begin{equation}
    \sigma_p\rho\geq\pi\hbar\left\{1+\frac{\rho^2}{4\pi^2}\left.\left[\frac{ R}{9}\left(\frac{N_aN^a}{N}-3\right)+\tilde{\xi}\nabla_a\left(N_b/\sqrt{N}\right)\nabla^a\left(N^b/\sqrt{N}\right)-\frac{N_a}{6\sqrt{N}}\Delta\left(N^a/\sqrt{N}\right)\right]\right|_{p_0}\right\}
\end{equation}
with the mathematical constant $\tilde{\xi}=\xi+7/6\pi^2\sim\Ord (10^{-1}),$ a result, which reflects the form of the ultrarelativistic momentum \eqref{ultrelmom}. In particular, it is independent of $\Pi.$ Hence, there is no divergence at high energies. Instead, the relation asymptotes towards a constant value. In principle, it therefore also applies to massless particles. On the other hand, the nonrelativistic shift-dependent correction in Eq. \eqref{nonrelunc} scales linearly with the mass of the particle. Thus, the gravitational influence is strongest on very massive particles as expected. Furthermore, as mentioned above, all relativistic corrections are dependent on the value of the shift vector. If the latter vanishes, the corrections only depend on the scalar curvature of the effective spatial geometry. It is instructive, to examine the new corrections by virtue of an example.

\subsection{Kerr black hole}

\begin{figure}
\begin{minipage}{.49\linewidth}
\centering
        \includegraphics[width=\linewidth]{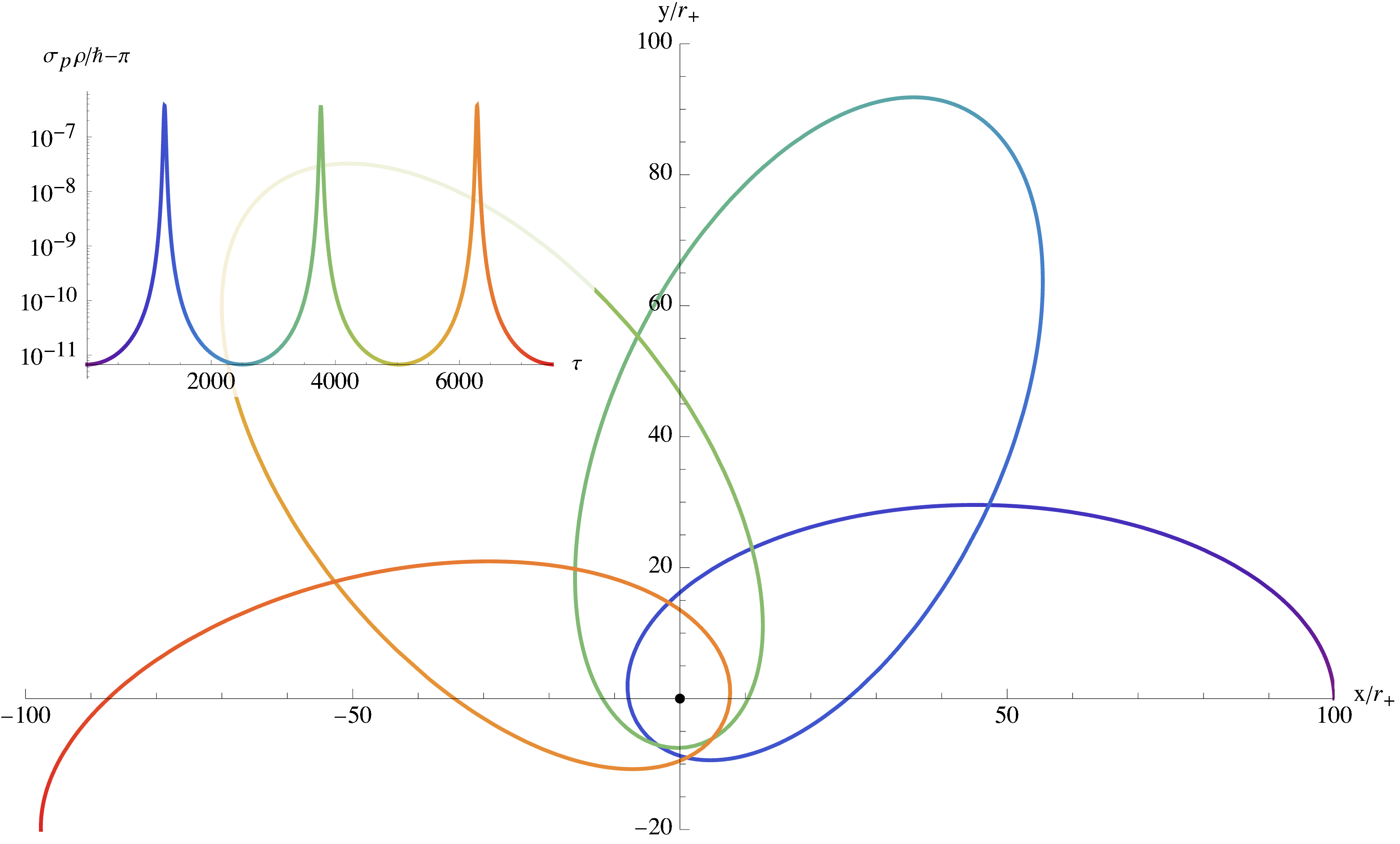}
\end{minipage}
\begin{minipage}{.49\linewidth}
\centering
        \includegraphics[width=.95\linewidth]{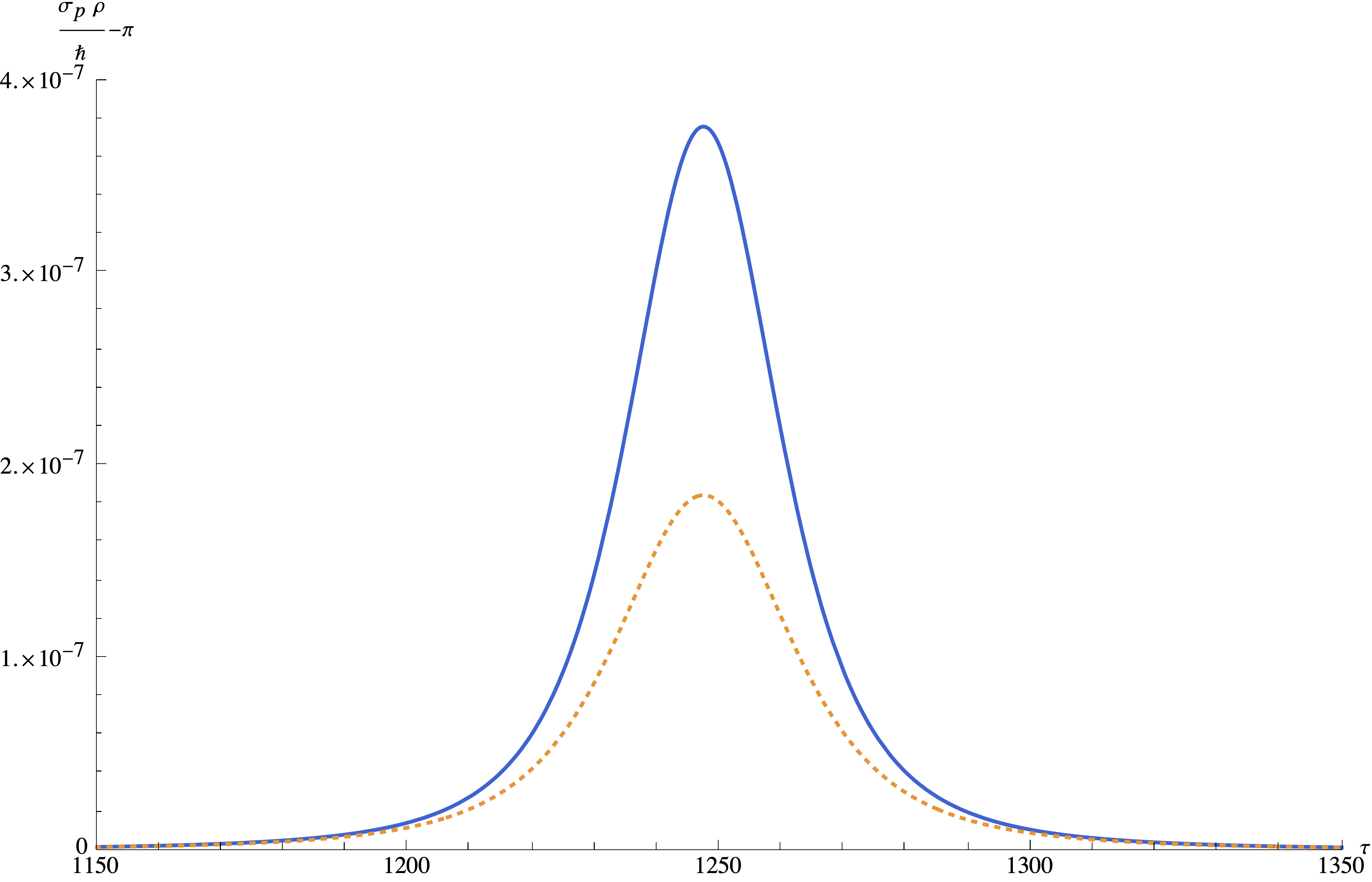}
\end{minipage}
\caption{The left plot shows the trajectory followed by a massive particle in the equatorial ($x$-$y$) plane of a fast black hole rotating as $a_J/GM=0.5$ and with an outer horizon of radius $r_+$ symbolized by the black disk in the center. Its starting point lies on the $x$ axis at a distance $100r_+$ from the source with initial velocity $u(\tau =0)\simeq(1.010,-0.0010,0.0000,0.0001)$ in Boyer-Lindquist coordinates. The color, ranging from violet to red, indicates an increase in the affine parameter $\tau.$ Inset in the top left corner is a plot displaying the all corrections to the uncertainty relation in units of $\hbar$ logarithmically experienced by a particle of mass $m=\hbar /r_+$ with position uncertainty $\rho=10^{-1}r_+$ along this orbit as a function of proper time. On the right-hand-side, the fully relativistic (blue) and nonrelativistic (orange, dotted) corrections are compared allowing a closer look at the first peak of the uncertainty relation.}
    \label{fig:relnonrelcomp}
\end{figure}

Arguably, the most famous geometry with nontrivial nondiagonal elements is the Kerr black hole, in Boyer-Lindquist coordinates described by the metric
\begin{align}
    \D s^2=&-\left(1+2\phi_{\text{GR}}\frac{r^2}{\Xi^2}\right)\D t^2+4\phi_{\text{GR}}a_J\sin^2\theta\frac{r^2}{\Xi^2}\D t\D \varphi\nonumber\\
    &+\frac{\Xi^2}{\Sigma}\D r^2+\Xi^2\D\theta^2+r^2\left(1+\frac{a_J^2}{r^2}-2\phi_{\text{GR}}\sin^2\theta\frac{a_J^2}{\Xi^2}\right)\sin^2\theta\D\varphi^2
\end{align}
with $\phi_{\text{GR}}=-GM/r$, $\Xi=r\sqrt{1+\cos^2\theta a_J^2/r^2}$ and $\Sigma=r^2(1+\phi_{GR}+a_J^2/r^2).$ The resulting relativistic uncertainty relation reads to fourth order in the gravitational potential $\phi_{\text{GR}}|_{r_0}$ and the relative angular momentum $a_J/r_0$ at the point $p_0=(r_0,\theta_0,\phi_0)$
\begin{equation}
    \sigma_P\rho\geq \pi\hbar\left\{1+\frac{\left.\phi^2_{\text{GR}}\right|_{r_0}\rho^2}{48\pi^2 r_0^2}\left[10+10\left.\phi_{\text{GR}}\right|_{r_0}+15\left.\phi^2_{\text{GR}}\right|_{r_0}-\frac{a_J^2}{r_0^2}\left(469-217\cos 2\theta-96\mathcal{F}_3|_{N=1}\left(7-3\cos 2\theta\right)\right)\right]\right\}.\label{KerrRes}
\end{equation}
Note, that, the radial coordinate, in terms of which this inequality is provided in Ref. \cite{Petruzziello21} is nonlinearly related to the one used in this paper. Therefore, even though the second and third terms in the square bracket have different prefactors, both results are equivalent. In said reference the modified uncertainty relation was evaluated along a wide trajectory around a rotating black hole for a heavy particle. The relativistic version clearly allows for closer orbits and lighter particles. Such an evolution is displayed in the left plot of Fig. \ref{fig:relnonrelcomp}, where the colour of the curve indicates progress in proper time. The graph inset into the top left corner of this visualisation provides the deviation of the right-hand side of inequality \eqref{KerrRes} from the value of the flat-space uncertainty $\hbar\pi$. Clearly, the influence is strongest, when the the curvature is large leading to peaks at the periapsis, as had already been concluded in Ref. \cite{Petruzziello21}. The nonrelativistic and relativistic expressions are compared graphically at the first peak in the plot to the right of Fig. \ref{fig:relnonrelcomp}. The former lead to an increase of the effect by a factor of two for the given choice of parameters.

\section{Conclusion}\label{sec:disc}

Starting at the dynamics of relativstic particles in curved spacetime, we have derived an uncertainty relation between the positions and momenta on hypersurfaces of constant time in accordance with the 3+1-formalism. This was achieved by, first, finding the relevant relativistic physical momentum operator and then obtaining its standard deviation on a compact domain. In particular, for reasons of simplicity we chose to work with geodesic balls whose radii are diffeomorphism invariant and yield, thus, a meaningful measure of position uncertainty. Under the assumption, that the involved position uncertainties are small in comparison to all relevant background curvature length scales, we solved the corresponding problem perturbatively by describing the effective spatial metric in terms of Riemann normal coordinates constructed in the center of the ball. This was done drawing heavily on results obtained in earlier work \cite{Dabrowski20,Petruzziello21}.

To second order, the resulting uncertainty relation is proportional to the Ricci scalar of the effective spatial metric as well as a couple of terms, which depend on the gradients of the shift vector and the lapse function. In particular, all relativistic corrections to the nonrelativistic result in Ref. \cite{Petruzziello21} are proportional to the shift vector and/ or its gradients. Thus, they all vanish in the absence of meaningful nondiagonal entries in the original spacetime metric. Interestingly, the ultrarelativistic limit asymptotically yields a constant correction and does not diverge. Thus, the relation may, in principle, also be used to describe massless particles. 

This formalism was applied to a particle moving on a geodesic in the equatorial plane of the Kerr-geometry. In that vein, it was shown, that the relativistic corrections support the effect by increasing the deviation from the flat-space uncertainty relation. 

Have we thus obtained a covariant uncertainty relation? Recall, that the derivation we followed throughout this paper was based on a given division of the underlying spacetime manifold $M$ into a time direction $\reals$ and spacelike hypersurfaces $\Sigma,$ on which the uncertainty relation is determined. To be more precise, this was indicated by the choice of domain. The investigated particle is confined to a geodesic ball at a certain time, which is clearly a slicing-dependent statement. Changing the time coordinate, \eg by a local Lorenz transformation, would lead to a deformation of the ball and change the problem entirely. Clearly, the core of this peculiarity lies in the fact that the objective lies in obtaining an uncertainty principle relating positions and momenta, notions, which can only be understood as absolute in an non-relativistic context. How could we, then, obtain an intrinsically covariant result?

In principle, the relevant quantity to study in this direction would be the standard deviation of the Dirac operator $\hatslashed{P}=-i\hbar\gamma^\mu\partial_\mu.$ By analogy, this requires a domain, compact not only in space but also in time, thus treating both entities equally. Hence, it seems necessary to consider the domain to be expanding into the future from an initial hypersurface to afterwards recollapse into another hypersurface, basically like the creation and subsequent anihilation of an excitation by the uncertainty. This will be the subject of future research.

\appendix

\section{Evaluation of $\left|\left|\text{MaxRe}\left[e^{i\Delta\phi}\Braket*{\psi_{n'l'm'}}{\hatslashed{\pi}^{(0)}\psi_{nlm}}\right]\right|\right|$}\label{app_maxre}

In order to be able to evaluate the expectation value of the momentum operator with respect to a general state $\Psi$ written in the basis of the Laplacian (see eq. \eqref{genstaten}), we need to compute the transition amplitudes $\Braket{\psi_{n'}}{\hatslashed{\pi}^{(0)}\psi_n}.$ In particular, confined to geodesic balls of radius $\rho$ and on a flat three-dimensional background they read
\begin{equation}
    \Braket*{\psi_{n'l'm'}}{\hatslashed{\pi}^{(0)}\psi_{nlm}}=\int_{B_\rho}\D\mu\psi_{n'l'm'}^*\hatslashed{\pi}^{(0)}\psi_{nlm},
\end{equation}
where the functions $\psi_{nlm}$ were defined in eq. \eqref{flatsol} and we introduced the flat space measure $\D\mu=\sigma^2\sin\chi\D\sigma\D\chi\D \gamma$ in spherical coordinates $\sigma^i=(\sigma,\chi,\gamma)$. 

According to eq. \eqref{vanmomrealwave}, those amplitudes evidently vanish if $n'=n,$ $l'=l$ and $m'=m.$ To be more precise, this result can be extended to cases where $m'\neq m.$ Having in mind that non-vanishing $\Delta m\equiv m'-m$ leads to a phase difference $\psi_{nlm}=\exp{i \Delta m\gamma}\psi_{nlm'},$ the only possible change in the transition amplitude has to stem from derivatives with respect to the coordinate $\gamma.$ Due to the proportionality
\begin{equation}
    \partial_\gamma\psi_{nlm}\propto \psi_{nlm},
\end{equation}
we infer, that the relevant integrals, \ie the ones which could prevent the transition amplitude from vanishing, share the behaviour
\begin{equation}
    \int_0^{2\pi} e^{-i\Delta m \gamma}\sin\gamma\D\gamma =\int_0^{2\pi} e^{-i\Delta m\gamma}\cos\gamma\D\gamma=0,
\end{equation}
where the last equality holds irrespective of the value of $\Delta m.$ Thus, varying solely $m$ does not change the transition amplitude, yielding
\begin{equation}
    \Braket*{\psi_{nlm'}}{\hatslashed{\pi}^{(0)}\psi_{nlm}}=0.
\end{equation}

As the eigenvalues of the Laplacian are functions of the quantum numbers $l$ and $m$ (\cf eq. \eqref{flateig}), the remaining transition amplitudes feature states with distinct eigenvalues. The evaluation of those can be simplified considering a different amplitude
\begin{equation}
    \Braket*{\psi_{n'l'm'}}{\hatslashed{\pi}_{(0)}^3\psi_{nlm}}=-\hbar^2\lambda_{nl}\braket*{\hatslashed{\pi}^{(0)}}=-\hbar^2\lambda_{n'l'}\braket*{\hatslashed{\pi}^{(0)}}-\hbar^3\int_{B_\rho}\D\mu\partial^j\left[\left(-i\slashed{\partial}\psi_{n'l'm'}\right)^*\partial_j \psi_{nlm}\right]\label{transappint},
\end{equation}
where the boundary condition \eqref{evp2} has been applied. The last term, being a total derivative, can be turned into a surface integral by Stokes' theorem such that we can rewrite eq. \eqref{transappint}
\begin{align}
    \braket*{\hatslashed{\pi}^{(0)}}=&\frac{\hbar}{\lambda_{nl}-\lambda_{n'l'}}\int_{\partial B_{\rho}}\D\tilde{\mu}\left(-i\slashed{\partial}\psi_{n'l'm'}\right)^*n^j\partial_j\psi_{nlm}\\
    =&\frac{\sigma^2\hbar}{\lambda_{nl}-\lambda_{n'l'}}\int_{S^2}\D\Omega\left(-i\slashed{\partial}\psi_{n'l'm'}\right)^*\partial_\sigma\psi_{nlm}\Big|_{\sigma=\rho}\label{surfint}
\end{align}
with the determinant of the induced metric on the surface of the geodesic ball, which in spherical coordinates is proportional the volume element of the two-sphere $S^2$ (of radius $\sigma$) $\D\tilde{\mu}=\sigma^2\D\Omega=\sigma^2\sin\chi\D\chi\D\gamma,$ and the outward normal $n^i=\delta^i_\sigma.$ Writing the basis states decomposed in terms of their radial and angular parts, 
\begin{equation}
R_{nl}(\sigma)=\sqrt{\frac{2}{\rho^3j^2_{l+1}(j_{l,n})}}j_{l}\left(j_{l,n}\frac{\sigma}{\rho}\right)    
\end{equation}
and the spherical harmonics $Y^l_m(\chi,\gamma)$ respectively, eq. \eqref{surfint} can be reexpressed as
\begin{equation}
    \Braket*{\psi_{n'l'm'}}{\hatslashed{\pi}^{(0)}\psi_{nlm}}=-\frac{i\rho^2\hbar}{\lambda_{nl}-\lambda_{n'l'}}\partial_{\sigma}R_{n'l'}\partial_{\sigma}R_{nl}\big|_{\sigma=\rho}\int\D\Omega \gamma_\sigma\left(Y^{l'}_{m'}\right)^*Y^l_m
\end{equation}
where $\gamma_\sigma=\gamma_a\partial\sigma^i/\partial x^a$ denotes the unit radial vector and we used the boundary condition \eqref{evp2} yielding $R_{nl}|_{\sigma=\rho}=0.$ Without loss of generality, we can choose $l\geq l'$ because the inverse case can be obtained from this one by complex conjugation. Then the remaining integral can be calculated explicitly yielding
\begin{align}
    \int\D\Omega\gamma_\sigma\left(Y^{l'}_{m'}\right)^*Y^l_m=\delta_{l'}^{l+1}\left[s^{-1}_{l'm'}\delta_{m'}^{m+1}\left( -\gamma_x+i\gamma_y\right)+s^{1}_{l'm'}\delta_{m'}^{m-1}\left( \gamma_x+i\gamma_y\right)+is^0_{l'm'}\delta^m_{m'}\gamma_z\right],
\end{align}
where we introduced the unit vectors in the $x,$ $y$ and $z$-directions denoted $\gamma_x,$ $\gamma_y$ and $\gamma_z,$ respectively, and the sequences $s^{\Delta m}_{lm}$
\begin{align}
    s^{\pm 1}_{lm}=&\frac{1}{2}\sqrt{\frac{(1+l\mp m)(2+l\mp m)}{3+4l(2+l)}},\\
    s^0_{lm}=&\sqrt{\frac{(l+1)^2-m^2}{4(l+1)^2-1}}.
\end{align}
Formulated explicitly, a general transition amplitude featuring the momentum operator reads
\begin{align}
    \Braket*{\psi_{n'l'm'}}{\hatslashed{\pi}^{(0)}\psi_{nlm}}=&\frac{\rho^2\hbar}{\lambda_{n'l'}-\lambda_{nl}}\partial_{\sigma}R_{n'l'}\partial_{\sigma}R_{nl}\big|_{\sigma=\rho}\Bigg\{\nonumber\\
    &\times\delta_{l'}^{l+1}\Bigg[s^1_{l'm'}\delta_{m'}^{m+1}\left(i \gamma_x+\gamma_y\right)
    +s^{-1}_{l'm'}\delta_{m'}^{m-1}\left(-i \gamma_x+\gamma_y\right)+is^0_{l'm'}\delta^m_{m'}\gamma_z\Bigg]\nonumber\\
    &+\delta_{l'+1}^{l}\Bigg[s^1_{lm}\delta_{m'+1}^{m}\left(-i \gamma_x+\gamma_y\right)+s^{-1}_{lm}\delta_{m'-1}^{m}\left(i \gamma_x+\gamma_y\right)-is^0_{l'm'}\delta^m_{m'}\gamma_z\Bigg]\Bigg\}.
\end{align}
Furthermore, the norm of its maximal real part including a relative phase $\Delta\phi$ becomes
\begin{align}
    \left|\left|\text{MaxRe}\left[e^{i\Delta\phi}\Braket*{\psi_{n'l'm'}}{\hatslashed{\pi}^{(0)}\psi_{nlm}}\right]\right|\right|=&\frac{\rho^2\hbar}{|\lambda_{n'l'}-\lambda_{nl}|}\partial_{\sigma}R_{n'l'}\partial_{\sigma}R_{nl}\big|_{\sigma=\rho}\nonumber\\
    &\times\left[\delta_{l'}^{l+1}\left(s^1_{l'm'}\delta^{m+1}_{m'}+s^{-1}_{l'm'}\delta^{m-1}_{m'}+s^0_{l'm'}\delta^m_{m'}\right)\right.\nonumber\\
    &+\left.\delta_{l'+1}^{l}\left(s^1_{lm}\delta^{m}_{m'+1}+s^{-1}_{lm}\delta^{m}_{m'-1}+s^0_{lm}\delta^m_{m'}\right)\right].\label{transnormmaxreal}
\end{align}
Introducing the sign function $\text{sgn}(x)$ which equals one for $x>0$ and negative one for $x<0,$ we can readily evaluate
\begin{equation}
    \partial_{\sigma}R_{nl}|_{\sigma=\rho}=\text{sgn}\left[j_{l+1}\left(j_{l,n}\right)\right]\sqrt{2\lambda_{nl}/\rho^3}
\end{equation}
and estimate $0<s^1_{lm},s^{-1}_{lm},s^0_{lm}< 1/2$ (though tighter estimates are possible). This implies for eq. \eqref{transnormmaxreal} that
\begin{align}
    \left|\left|\text{MaxRe}\left[e^{i\phi}\Braket*{\psi_{n'l'm'}}{\hatslashed{\pi}^{(0)}\psi_{nlm}}\right]\right|\right|\leq & \frac{\hbar}{\rho}\frac{\sqrt{\lambda_{nl}\lambda_{n'l'}}}{|\lambda_{n'l'}-\lambda_{nl}|}\nonumber\\
    &\times \left[\delta^{l'+1}_{l}\left(\delta^{m+1}_{m'}+\delta^{m-1}_{m'}+\delta^m_{m'}\right)+\delta^{l'}_{l+1}\left(\delta^{m}_{m'+1}+\delta^{m}_{m'-1}+\delta^m_{m'}\right)\right].\label{res_app_maxre}
\end{align}

\bibliographystyle{unsrt}
\bibliography{bib}

\end{document}